\documentclass[10pt]{article} 
\pdfoutput=1
\usepackage{amssymb,amsmath,amsfonts,theorem,subfigure} 
\usepackage{graphicx}
\graphicspath{{./pdf/}}
\usepackage{euler,palatino}
\usepackage{pstricks}
\numberwithin{equation}{section}

\newcommand{\be}{\begin{equation}}
\newcommand{\ee}{\end{equation}}

\title{{\huge Geometric Exponents, SLE\\[4pt] and Logarithmic Minimal Models}}
\author{
{\Large Yvan Saint-Aubin}\footnote{\ttfamily saint{\char'100}dms.umontreal.ca}\\[6pt]
\it D\'epartement de math\'ematiques et de statistique\\ 
\it Universit\'e de Montr\'eal, C.P.\ 6128, succ.\ centre-ville, Montr\'eal\\ 
\it Qu\'ebec, Canada, H3C 3J7\\[10pt]
{\Large Paul A. Pearce}\footnote{\ttfamily P.Pearce{\char'100}ms.unimelb.edu.au}, \quad
{\Large J\o{}rgen Rasmussen}\footnote{\ttfamily J.Rasmussen{\char'100}ms.unimelb.edu.au}\\[6pt]
\it Department of Mathematics and Statistics\\ 
\it University of Melbourne\\
\it Parkville, Victoria 3010, Australia}
\date{\empty}

\begin{document} 
\maketitle

%
%
 
\begin{abstract}
\noindent
In statistical mechanics, observables are usually related to local degrees of freedom 
such as the $Q<4$ distinct states of the $Q$-state Potts models or the heights of the restricted 
solid-on-solid models. In the continuum scaling limit, these models are described by rational 
conformal field theories, namely the minimal models ${\cal M}(p,p')$ for suitable $p,p'$. 
More generally, as in stochastic Loewner evolution (SLE$_\kappa$), one can consider 
observables related to nonlocal degrees of freedom such as paths or boundaries of clusters. 
This leads to fractal dimensions or geometric exponents related to values of conformal 
dimensions not found among the finite sets of values allowed by the rational minimal models. 
Working in the context of a loop gas with loop fugacity $\beta=-2\cos\frac{4\pi}{\kappa}$, we use 
Monte Carlo simulations to measure the fractal dimensions of various geometric objects such 
as paths and the generalizations of cluster mass, cluster hull, external perimeter and red bonds. 
Specializing to the case where the SLE parameter $\kappa=\frac{4p'}{p}$ is rational with $p<p'$, 
we argue that the geometric exponents are related to conformal dimensions found in the 
infinitely extended Kac tables of the logarithmic minimal models ${\cal LM}(p,p')$. 
These theories describe lattice systems with 
nonlocal degrees of freedom. 
We present results for critical dense polymers ${\cal LM}(1,2)$, critical percolation 
${\cal LM}(2,3)$, the logarithmic Ising model ${\cal LM}(3,4)$, the logarithmic tricritical 
Ising model ${\cal LM}(4,5)$ as well as ${\cal LM}(3,5)$.
Our results are compared with rigourous results from SLE$_\kappa$, with predictions 
from theoretical physics and with other numerical experiments.
Throughout, we emphasize the relationships between SLE$_\kappa$, geometric 
exponents and the conformal dimensions of the underlying CFTs.\\[0pt]

\end{abstract}

%
%

\newpage
\tableofcontents

%
%


\section{Introduction}
\label{sec:intro}

The numerical measurements of fractal dimensions and geometric exponents presented 
here aim at strengthening the ties between three central para\-digms describing 
two-dimensional phase transitions. The first, the lattice approach, which started with 
Onsager's calculation~\cite{Onsager} of the spectrum of the transfer matrix of the Ising 
model and continued in the work of Baxter~\cite{BaxBook}, is the closest to a first-principle 
description of phase transitions. Rigorous results obtained from this approach are important, 
but remain far from the more physical continuum field-theoretic description used in solid-state 
or elementary particle physics. By limiting the description to the critical point, the second 
paradigm, conformal field theory (CFT)~\cite{FMS}, has provided such a full-fledged 
continuum field theory with exact calculations of certain correlation functions. In doing so, 
it enlarges the set of algebraic tools, already central in Onsager's solution, that describe 
critical behaviour. Even though CFT enables perturbation away from criticality, the approach 
to criticality is far from being understood. Moreover, its description of statistical models 
misses the probability framework that one would expect from models whose ab initio description 
uses the Boltzmann distribution. The third paradigm, Stochastic Loewner Evolution 
(SLE)~\cite{Schramm,Lawler,Werner,Cardy05,KagerNienhuis03,BauerBernard,Duplantier}, 
is indeed probabilistic in nature. 
In SLE, objects are of a measure-theoretic nature and are concrete enough to reproduce rigorously several predictions from the other two approaches, mostly those from CFT.
Since rigorous derivations~\cite{Beffara04} give the fractal dimension of certain SLE paths confirming predictions from CFT, it seems natural to us that the study of fractal properties of critical geometric objects will improve the understanding of the ties between the three paradigms of lattice statistics, CFT and SLE.

Many geometric objects have been introduced to probe the critical behaviour of lattice models: 
the hull of a spin cluster, its mass, its external perimeter and its red bonds. 
In contrast to the usual correlation functions of local degrees of freedom, these geometric 
observables relate to nonlocal objects on the lattice. Theoretical predictions for various 
fractal dimensions and geometric scaling exponents were made in the late 
eighties~\cite{SaleurDuplantier}. Many numerical measurements have been made, 
but only relatively recently were good 
measurements~\cite{AAMRH,JankeSchakel05d,JankeSchakel06b} 
made of the fractal dimensions of these objects.  
These support the theoretical predictions but are limited to the Fortuin-Kasteleyn (FK) 
clusters~\cite{FortuinKasteleyn} of $Q$-state Potts models~\cite{Potts}. 

The basic random variables in lattice spin models used for Monte Carlo (MC) simulations are the Potts spins sitting at the vertices of the lattice. Reasons to use Potts spin models for these simulations are numerous. First, spin variables are close to physical objects, readily interpretable as originating from an atom in a crystal. Second, the Swendsen-Wang algorithm~\cite{SwendsenW} provides extremely quick upgrades in 
MC simulations. Third, the thermodynamic limits of three of the $Q$-Potts models are archetypal lattice statistical models, namely, critical percolation ($Q\to 1$) with central charge $c=0$, the Ising model ($Q=2$) with $c=\frac12$ and the $3$-state Potts model with $c=\frac 45$. 

The continuum scaling limits of the $Q=2,3$ Potts models are described by rational CFTs. More explicitly, they are described by the members ${\cal M}(3,4)$ and ${\cal M}(5,6)$ of the series of minimal models ${\cal M}(p,p')$~\cite{BPZ,RSOS} with central charges 
\be
  c=c(\bar\kappa)=1-\frac{6(1-\bar\kappa)^2}{\bar\kappa}
    =13-6\big(\bar\kappa+\frac{1}{\bar\kappa}\big),
    \qquad \bar\kappa=\frac{p'}{p}>1,\qquad\mbox{$p,p'>1$ coprime}
\label{cM}
\ee
related to $Q$ by the mapping
\be
  \sqrt{Q}=-2\cos\frac{\pi}{\bar\kappa},\qquad Q<4.
\label{PottsMapping}
\ee
But now we are immediately presented with two problems. First, critical percolation, which formally 
is an allowed minimal model corresponding to $p=2$, $p'=3$, is {\em empty} in the sense that it 
only contains the identity with conformal dimension $\Delta=0$. Second and more generally, 
the conformal dimensions related to the nonlocal geometric observables are not 
found~\cite{SaleurDuplantier}
among the admissible conformal dimensions in the Kac tables for these rational CFTs.

To circumvent these problems in considering fractal geometric objects, we follow \cite{GuoBlote06,Cardy06} and move the lattice context away from the rational $Q$-state Potts or Restricted Solid-On-Solid (RSOS) models~\cite{RSOS}
and instead consider a loop gas with loop fugacity
\be
  \beta=-2\cos\frac{\pi}{\bar\kappa}.
\label{beta}
\ee
We believe there is good ground to measure the fractal dimensions of suitable geometric objects directly in the loop gas. In the loop gas, the basic random variables
are the nonlocal degrees of freedom associated with segments of the loops. The continuum scaling limit of such a loop gas is not described by a rational CFT --- it is described by the 
Coulomb gas~\cite{Coulomb,FMS}. 
Moreover, if $\bar\kappa$ is rational, which is the case of prime interest here, the loop gas is equivalent to the logarithmic minimal model 
${\cal LM}(p,p')$~\cite{PRZ} and exhibits some remarkable properties. 
In the continuum scaling limit, it is thus described by a {\em logarithmic} CFT~\cite{LogCFT}
with central charge given by the same expression as in the rational case (\ref{cM})
\be
 c=1-\frac{6(1-\bar\kappa)^2}{\bar\kappa}
   =13-6\big(\bar\kappa+\frac{1}{\bar\kappa}\big),\qquad 
   \bar\kappa=\frac{p'}{p}>1,\qquad\mbox{$p,p'\ge 1$ coprime}.
\label{cLM}
\ee
It is noted, though, that the logarithmic minimal model ${\cal LM}(p,p')$ is defined also for 
$p=1$.
Critical dense polymers, with $c=-2$, now appears as the member ${\cal LM}(1,2)$
and critical percolation as the member ${\cal LM}(2,3)$.
For recent studies of critical percolation as a logarithmic CFT, we refer to~\cite{MR0708} and
references therein.
As we will argue in many cases, 
the conformal dimensions related to the observed geometric exponents are found among 
the conformal dimensions in the {\em infinitely} extended Kac tables (\ref{logConformalDims}) for these logarithmic CFTs.

The basic probabilistic object of chordal SLE$_\kappa$ is a path or curve with (at least) its starting point 
on the boundary of the domain within which it evolves. The evolution is described by (\ref{SLE}) 
where we identify 
\be
  \kappa=4\bar\kappa.
\ee
Probabilists working with SLE use $\kappa=6$ for critical percolation while physicists working 
with CFT use $\bar\kappa=\frac32$.
One can argue that studying a lattice loop gas is to probe directly a discretized version of SLE 
paths. Alternatively, it has been proved for some values of $\kappa$
that the curves formed by loop segments of the loop gas 
coincide, in the continuum scaling limit, with SLE paths. Throughout this paper, we will 
emphasize the relationships between SLE, geometric exponents and the conformal 
dimensions of the underlying CFTs. As already indicated, for rational values of $\bar\kappa=p'/p$, 
we argue that the relevant underlying CFT is that of the logarithmic minimal model ${\cal LM}(p,p')$.

The paper is organized as follows. The next section recalls the definition of the SLE process 
and summarizes the theoretical predictions for the fractal dimensions and geometric 
exponents of interest. Section~\ref{sec:defect} is devoted to the measurement of the fractal 
dimension of a path associated with a defect. This is the most natural discretization of an 
SLE path. Section~\ref{sec:HMEPRB} introduces natural extensions of classical geometric 
objects (cluster hull, cluster mass, external perimeter and red bonds) to the context of a loop 
gas. The measurement of these quantities by MC methods is also presented in this section. 
The lattice loop gas and the logarithmic minimal models ${\cal LM}(p,p')$ are recalled 
in Appendix~A.  The various MC upgrade algorithms that we use are discussed in Appendix~B. 
Finally, in Appendix~C, we describe the initial thermalization of the MC process as well as 
checks on the statistical independence of our measurements.

\section{SLE and CFT}

\subsection{SLE}

Two landmark results led to the development of SLE as a tool to probe critical phenomena.
One of them is Smirnov's proof~\cite{Smirnov} that one observable of critical percolation
on the triangular lattice has a scaling limit that is conformally invariant. The second
is the introduction by Schramm~\cite{Schramm} of a one-parameter family of conformally invariant measures
on curves in the half-plane, denoted by SLE$_\kappa$. Schramm also showed that this
family exhausts all such conformally invariant measures. Hence, the measure on interfaces
of percolating clusters must fall in this family. Schramm showed that SLE$_6$ was the one.

A simple way to understand this measure it to see the interfaces (or curves) as being grown
out of a point on the real line, the boundary of the half-plane. During the growth, the interface
may touch itself or the real line. Points belonging to the interface and points from which
a curve cannot be drawn to infinity without crossing the interface form the SLE hull. 

Let us denote by $\mathcal C_t$ the complement of the hull at time $t$.
Then $\mathcal C_t\cap \mathbb H$  is a simply-connected
open set and, by the Riemann mapping theorem, there exists an analytic function 
$g_t:\mathcal C_t \cap \mathbb H\rightarrow \mathbb H$ that is one-to-one. 
It is this map $g_t$ that satisfies the stochastic Loewner equation:
\be
 \partial_t g_t(z)=\frac{2}{g_t(z)-\sqrt{\kappa}\,B_t},\qquad t>0, \quad \text{\rm with\ }g_0(z)=z,
\label{SLE}
\ee
where $B_t$ is a one-dimensional Brownian motion starting at $0$. To determine uniquely $g_t$,
its behaviour  around infinity is fixed by $\lim_{z\rightarrow\infty}(g_t(z)-z)=0$. The
interface $\gamma=\gamma_t$ is then defined as the (continuous) path $\gamma:[0,\infty)
\rightarrow \bar{\mathbb H}$ with $g_t(\gamma_t)=\sqrt\kappa B_t$. Equation \eqref{SLE}
determines (somewhat implicitly) the probabilistic properties of $\gamma_t$ in terms of those
of the Brownian motion.

One of the important corollaries of stochastic Loewner evolution is that the curve 
$\gamma$ generated by the process with parameter $\kappa$ has Hausdorff dimension 
equal to 
\begin{equation}
 d_{\text{\rm path}}^{\text{\rm SLE}}=\min(2,1+{\textstyle{\frac{\kappa}8}})
\label{Beff}
\end{equation} 
almost surely. 
This result, due to Beffara~\cite{Beffara04}, 
together with the fact that the scaling limit of percolation is SLE$_6$,
proves one of the important predictions of theoretical physics, at least for 
one model on a very particular lattice. If one believes in universality, then the result should 
also apply to other lattices. 
Note that, if one sets $\kappa=0$ in \eqref{SLE}, the measure degenerates and the interface 
$\gamma_t$
tracing the imaginary axis has probability one. The fractal dimension of $\gamma$ is then $1$.
For $\kappa$ larger or equal to $8$, the curve is space filling and has dimension $2$.

\subsection{Logarithmic minimal models and CFT}

The logarithmic minimal models ${\cal LM}(p,p')$ \cite{PRZ}
are reviewed briefly in Appendix~\ref{AppLoop}. 
The degrees of freedom of these theories are nonlocal objects associated with segments of the loops. 
In the continuum scaling limit, these models are described by logarithmic CFTs with central charges
(\ref{cLM}).
Conformal dimensions of these CFTs can be organized into infinitely extended Kac tables 
\be
  \Delta_{r,s}=\frac{(\bar\kappa r-s)^2-(\bar\kappa-1)^2}{4\bar\kappa},\qquad r,s\in\mathbb{N}.
\label{logConformalDims}
\ee
There is no claim here, however, that these values exhaust all possible values of the conformal dimensions for these theories. Indeed, there is some evidence~\cite{SaleurDuplantier} 
that conformal dimensions corresponding to half-integer values of $r$ and $s$ are allowed. 

There is a duality in these theories, mimicked in $SLE_\kappa$ and described by
\be
  p\leftrightarrow p',\quad\text{\rm or}\quad\kappa \leftrightarrow \frac{16}{\kappa},
   \quad\text{\rm or}\quad \bar\kappa\leftrightarrow\frac1{\bar\kappa}. 
\ee
Under this duality, the central charges and conformal dimensions verify 
\be
  c(\bar\kappa)=c\left({\textstyle{\frac1{\bar\kappa}}}\right),\qquad
  \Delta_{r,s}\big(\frac{1}{\bar\kappa}\big)=\Delta_{s,r}(\bar\kappa).
\ee
Here we only consider the case $p<p'$ with $\bar\kappa=\frac {p'}p>1$. From the lattice, 
the dual models with $p>p'$ and $\bar\kappa<1$ are expected to come from the dilute 
logarithmic minimal models with central charge 
$c(\bar\kappa)=c\left({\textstyle{\frac1{\bar\kappa}}}\right)$ but we do not consider these models here.

\subsection{Relation between geometric exponents and conformal dimensions}

An important challenge is to relate the values of various geometric exponents or fractal dimensions $d$ of observables to the conformal dimensions $\Delta$ of correlators
\begin{equation}
 \langle \phi(0)\phi(x)\rangle \sim |x|^{-2\Delta}
\end{equation}
in the underlying CFTs, for example the critical $Q$-state Potts models.
For $Q$ integer, these are rational CFTs with relatively few conformal dimensions and sometimes one is led to conformal dimensions not found among the admissible values for these 
rational theories. 
Some observables in the logarithmic minimal models ${\cal LM}(p,p')$ should correspond to observables in the Potts models related to FK clusters. 
In addition, however, the logarithmic minimal models ${\cal LM}(p,p')$ allow us to explain many  
observables {\em not} explained by the rational models. 
Consider a geometric observable ${\cal O}$, measured on FK clusters 
of linear scale or radius $R$ and described by correlators of the field $\phi$. An example is $C$, the cluster mass (total number of sites) of a percolating cluster, or $H$, its hull. 
A heuristic scaling argument in two dimensions then gives
\begin{equation}
 {\cal O}(R) \sim \int d^2x\, \langle \phi(0)\phi(x)\rangle
  \sim \int d^2x\, |x|^{-2\Delta} \sim R^{2-2\Delta} \sim R^{d_{\cal O}}
\end{equation}
where $d_{\mathcal O}$ stands for the fractal dimension of the observable $\mathcal O$.
In this way, such observables are associated with particular conformal weights in the Kac table. The renormalization group or geometric exponent $y$ is related to $\Delta$ by
\begin{equation}
 y=2-2\Delta=d_{\cal O}.
\end{equation}

In Section~\ref{sec:defect}, we verify, through MC sim ulations, that precisely the fractal dimensions 
given in (\ref{Beff}) are found in the continuum scaling limit of the logarithmic minimal models 
${\cal LM}(p,p')$ with boundary conditions imposing a single (chordal) defect propagating through 
the system. 
We find it natural to identify this defect with (a discretized version of) the SLE path.
We also note that for all $p,p'$, 
the dimension $d_{\text{\rm path}}^{\text{\rm SLE}}$ given above corresponds to
\begin{equation}
 d_{\text{\rm path}}^{\text{\rm SLE}}=2-2\Delta_{p,p'-1}=y_{p,p'-1}
\end{equation}
where $\Delta_{p,p'-1}$ is a conformal dimension (\ref{logConformalDims}) of the logarithmic minimal model that falls outside the rational Kac table.

In Section~\ref{sec:otherMasses}, we recall the definitions of the hull, cluster mass, 
external perimeter and red bonds for the context of the FK clusters of the $Q$-state Potts 
models and show how these definitions can be extended to logarithmic minimal models 
${\cal LM}(p,p')$. 
The fractal dimensions (cluster mass) of FK clusters are given by (see~\cite{Duplantier,JankeSchakel06b} and references therein)
\begin{equation}
 {\textstyle{\Delta_C^{FK}=\Delta_\sigma=\frac{1}{2}-\frac{3\bar\kappa}{16}-\frac{1}{4\bar\kappa},
   \qquad  d_C^{FK}=1+\frac{3\bar\kappa}{8}+\frac{1}{2\bar\kappa}}}
\label{CFK}
\end{equation}
with the identification 
\begin{equation}
  C=\{\mbox{Cluster mass}\}\leftrightarrow \Delta_\sigma=\Delta_{(p\pm 1)/2,p'/2}.
\end{equation}
Notice that, depending on the parities of $p$ and $p'$, the Kac labels here can take half-integer 
values as foreshadowed after (\ref{logConformalDims}). 
We also have the following associations (see~\cite{Duplantier, JankeWeigel06} and references therein)
\begin{equation}
 H=\{\mbox{Hull}\}\leftrightarrow\Delta_{p,p'\pm 1},\quad 
 EP=\{\mbox{External Perimeter}\}\leftrightarrow\Delta_{p\pm 1,p'},\quad 
 RB=\{\mbox{Red Bonds}\}\leftrightarrow\Delta_{p,p'\pm 2}
\end{equation}
with conformal and fractal dimensions
\begin{alignat}{4}
 \Delta_{H}^{FK}\;&=&\ \Delta_{p,p'\pm 1}&={\textstyle{\frac{1}{2}-\frac{\bar\kappa}{4}}},&
   \qquad\qquad   d_H^{FK}\;&={\textstyle{1+\frac{\bar\kappa}{2}}}\\
  \Delta_{EP}^{FK}&=&\ \Delta_{p\pm 1,p'}&={\textstyle{\frac{1}{2}-\frac{1}{4 \bar\kappa}}},&
  \qquad\qquad   d_{EP}^{FK}&={\textstyle{1+\frac{1}{2\bar\kappa}}}\\
  \Delta_{RB}^{FK}&=&\ \Delta_{p,p'\pm 2}&={\textstyle{\frac{1}{2}-\frac{\bar\kappa}{4}
    +\frac{3}{4 \bar\kappa}}},&\qquad\qquad  d_{RB}^{FK}&
   ={\textstyle{1+\frac{\bar\kappa}{2}-\frac{3}{2\bar\kappa}}}
\end{alignat}
We summarize in Table~\ref{tab:allFractalDimensions} the values of the fractal dimensions 
of these observables for the logarithmic minimal models ${\cal LM}(p,p')$ that will be studied. 
In Section~\ref{sec:HMEPRB}, we verify, through MC simulations, that precisely these fractal 
dimensions are found by looking 
at suitably defined nonlocal 
observables of these logarithmic minimal models ${\cal LM}(p,p')$.
We note that the fractal dimension of the cluster hull $d_H$ coincides with the fractal dimension
of the SLE path $d_{\text{\rm path}}^{\text{\rm SLE}}$. 

\begin{table}[htbp]
\bigskip
\begin{center}\leavevmode
\begin{tabular}{|c||c|c|c|c||c|c|c|c|}
\hline
&&&& &&&& \\[-6pt]
 model & $\kappa$ & $\bar\kappa$ & $c$ & $\beta=\sqrt{Q}$ & $d_H$ & $d_C$ & $d_{EP}$ 
  & $d_{RB}$ \\
&&&& &&&& \\[-6pt]
\hline\hline
&&&& &&&& \\[-6pt]
  ${\cal LM}(p,p')$ & $\frac{4p'}{p}$ & $\frac{p'}{p}$ & $13-6\bar\kappa-\frac{6}{\bar\kappa}$ & 
  $-2\cos\frac{\pi}{\bar\kappa}$ & $1+\frac{\bar\kappa}{2}$ & $1+\frac{3\bar\kappa}{8}
  +\frac{1}{2\bar\kappa}$ & $1+\frac{1}{2\bar\kappa}$ & $1+\frac{\bar\kappa}{2}-\frac{3}{2\bar\kappa}$  \\
&&&& &&&& \\[-6pt]
\hline
&&&& &&&& \\
${\cal LM}(1,2)$ & $8$ & $2$ & $-2$ & $0$ & $2$ & $2$ & $\frac{5}{4}$ & $\frac{5}{4}$  \\
&&&& &&&&\\
 ${\cal LM}(3,5)$ & $\frac{20}{3}$ & $\frac{5}{3}$ & $-\frac{3}{5}$ & $\frac{1}{2} (\sqrt{5}-1)$ & $\frac{11}{6}$ &$\frac{77}{40}$ & $\frac{13}{10}$ & $\frac{14}{15}$ \\
&&&& &&&& \\
${\cal LM}(2,3)$ & $6$ & $\frac{3}{2}$& $0$ & $1$ & $\frac{7}{4}$ & $\frac{91}{48}$ & $\frac{4}{3}$ & $\frac{3}{4}$ \\
&&&& &&&& \\
${\cal LM}(3,4)$ & $\frac{16}{3}$ & $\frac{4}{3}$& $\frac{1}{2}$ & $\sqrt{2}$ & $\frac{5}{3}$ & $\frac{15}{8}$ & $\frac{11}{8}$ &$\frac{13}{24}$ \\
&&&& &&&& \\
${\cal LM}(4,5)$ & $5$ & $\frac{5}{4}$ & $\frac{7}{10}$ & $\frac{1}{2} (\sqrt{5}+1)$ & $\frac{13}{8}$ & $\frac{299}{160}$ & $\frac{7}{5}$ & $\frac{17}{40}$ \\
&&&& &&&& \\
\hline
\end{tabular}
\end{center}
\caption{The models, their parameters $\kappa$, $\bar\kappa$, $c$, $\beta$ and fractal dimensions. 
The models include critical dense polymers ${\cal LM}(1,2)$, critical percolation ${\cal LM}(2,3)$, 
the logarithmic Ising model ${\cal LM}(3,4)$ and the logarithmic tricritical Ising model ${\cal LM}(4,5)$.
The notations $H, C, EP, RB$ stand for Hull, Cluster mass, External Perimeter and Red Bonds. 
The fractal dimension of the hull coincides with the fractal dimension of the SLE path 
$d_{\text{\rm path}}^{\text{\rm SLE}}$ as given in (\ref{Beff}). 
The fractal dimensions of the cluster, external perimeter and red bonds coincide with those 
of the FK clusters in the associated $Q$-state Potts models via (\ref{PottsMapping}).}
\label{tab:allFractalDimensions}
\end{table}

\section{Fractal dimension of a defect}
\label{sec:defect}

\subsection{Definition and measurements}
\label{sec:defMeasure}

For the simulations, we use the logarithmic minimal model ${\cal LM}(p,p')$, a loop 
gas introduced in~\cite{PRZ}. 
Some very rudimentary properties of these models are
recalled in Appendix~\ref{AppLoop}. 
When the parameter $\bar\kappa$ characterizing the model is chosen to describe 
one of the $Q$-Potts models, the distribution of the loops coincides with that of the FK contours. 
We refer to an open curve in a loop configuration, that is, one starting and ending on the 
boundary of the domain considered, as a {\em defect}. Because of the parallel between loops 
in the logarithmic minimal models and interfaces of FK graphs, it is natural to consider a defect 
as a (discretized version of a) chordal SLE process. The goal of the present section is to measure 
the fractal dimension of such a defect for 
models corresponding to various values of 
$\bar\kappa$ of the SLE process. 
In the case of critical dense polymers ${\cal LM}(1,2)$ where $\bar\kappa=2$, the loop fugacity
vanishes, $\beta=0$, hence excluding all configurations with at least one closed loop.
In this model, the defect fills the space and the agreement with 
$d_{\text{\rm path}}^{\text{\rm SLE}}=2$ is trivial. 
This model is not studied further in this section.

\begin{figure}[h!]
\begin{center}\leavevmode
\includegraphics[width = 0.4\textwidth]{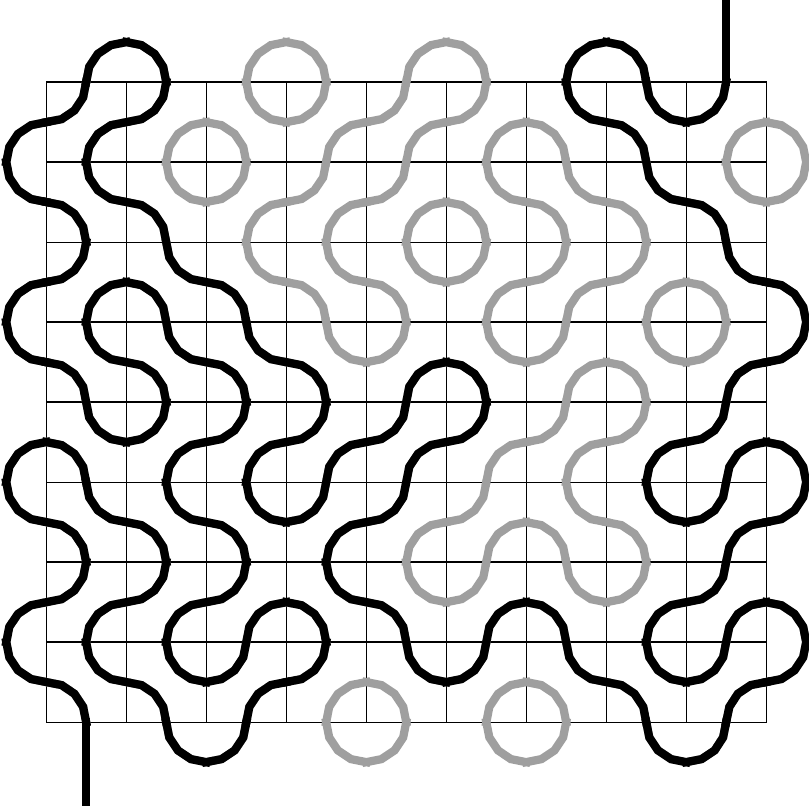}
\end{center}
\caption{A configuration on a $H\times V=9\times 8$ lattice. 
The defect visits $N_\Delta=104$ quarter-circles, including the ones on the boundary.}
\label{fig:8x9}
\end{figure}

There are various definitions of dimension for objects of fractal nature. The Hausdorff dimension 
is widely used in analysis, but it is customary to use the Minkowski definition for simulations. 
Let $D$ be a set $D\subset \mathbb R^d$ and let $N(\epsilon)$ be the number of boxes 
of side length $\epsilon$ required to cover $D$ in an evenly-spaced grid. 
The Minkowski dimension of $D$ is then given by
\begin{equation}
  d_D=\lim_{\epsilon\rightarrow 0}\frac{\log N(\epsilon)}{\log(1/\epsilon)}
\label{eq:minkowski}
\end{equation}
if the limit exists. There are sets whose Hausdorff and Minkowski dimensions differ, but this subtlety 
is somewhat irrelevant here since we are to approximate the measurement of fractal objects by 
studying {\em finite} geometric objects. 

The objective here is to examine, using simulations, whether the fractal dimension of
a defect $d_D$ indeed corresponds to the fractal dimension of the SLE path 
$d_{\text{\rm path}}^{\text{\rm SLE}}$
\be
 d_D\overset{?}{=}d_{\text{\rm path}}^{\text{\rm SLE}}.
\ee
However, since we work on a finite lattice, say of size $H\times V$, we want to consider a 
discretization
$d_D^{H\times V}$ of the fractal dimension defined on the {\em finite} lattice
yet approximating $d_D$ in the infinite limit.
One may think of the object $D$ (the defect) as being drawn on a rectangle whose size and 
aspect ratio are fixed; the increase in $H$ and $V$, while keeping their ratio fixed, then amounts 
to the limiting process. Therefore, if the size of the rectangle is fixed, $H\times V$ is related 
to the side length $\epsilon$ of a box as $H\times V\sim \frac1{\epsilon^2}$ or
\be
 \log (1/\epsilon)\sim \frac12\log (\# \text{\rm\ of boxes}).
\ee
In fact, as each box can be crossed twice, one should instead count the number of quarter-circles,
that is, the number of half-boxes. We therefore consider
\begin{equation}
 d_D^{H\times V}\sim 2\times \frac{\log N_\triangle(D)}{\log(2H\times V)}
  \sim\frac{\log N_\triangle(D)}{\log(R)} 
\label{eq:minkowski2}
\end{equation}
where $N_\triangle(D)$ is the number of quarter-circles the defect $D$ visits and where 
we have introduced some linear scale $R$. Up to a linear factor, $R$ should be approximately 
$(2H\times V)^\frac12$. If the defect crosses the same box twice, this counts for $2$ in $N_\triangle(D)$. 
Figure~\ref{fig:8x9} depicts the setting that we use for measurements 
of $d_D^{H\times V}$. The lattice is chosen to be almost square as the number $H$ of 
boxes in a row is equal to $V+1$ where $V$ is the number of boxes in a column. 
We take $V$ to be even. 
This allows for all loops on vertical boundaries to be closed by half-circles, as depicted in the
figure. Because $H=V+1$, all loops but one reaching a horizontal boundary can also be closed 
by half-circles. If we choose the pair of loose ends at the top right and 
bottom left, respectively, the {\em defect} will cross the lattice and be a macroscopic object. 
It is for this geometric object that $N_\Delta$ is computed for each configuration of the sample. 
To probe the asymptotic limit $d_D$, we will have to repeat these measurements on several 
lattices, starting with very small ones $H\times V=3\times 2$ and proceeding to 
$513\times 512$ or $1025\times 1024$, depending on the model.

\begin{figure}[h!]
\begin{center}\leavevmode
\includegraphics[width = 0.99\textwidth]{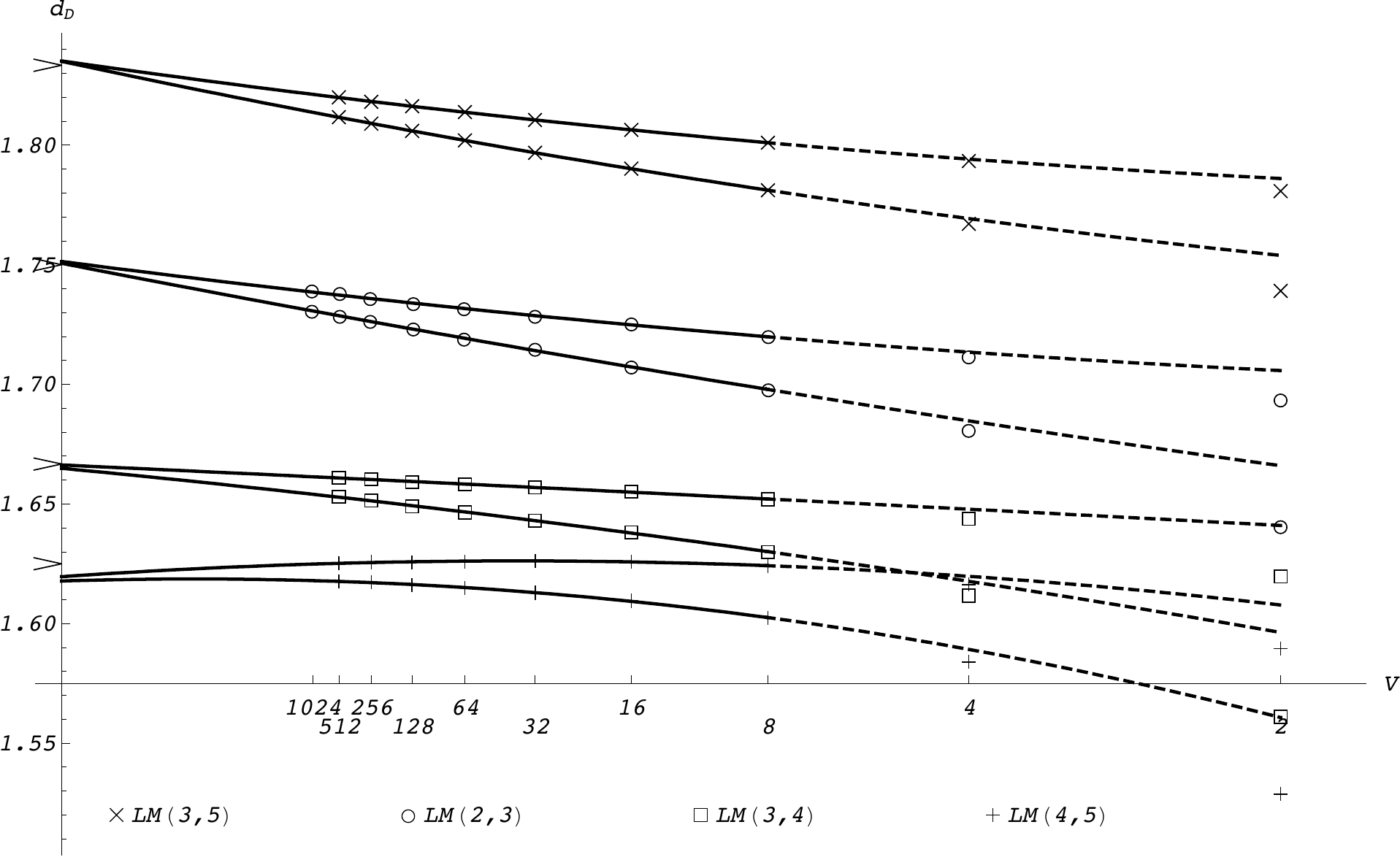}
\end{center}
\caption{The data $\{d_{D,1}^{H\times V}\}$ (lower curve) and $\{d_{D,2}^{H\times V}\}$ 
(upper curve) for the four models studied. The curves are the fits obtained as described in 
Section~\ref{sec:extract}, to which end only lattices with $H\times V\ge 9\times 8$ are used.
Here and throughout the paper, the error bars are much smaller than the symbols
used to depict the data.
}
\label{fig:dimDefect}
\end{figure}

The technical details of the statistical analysis are given in Section~\ref{sec:TI}. 
Here we present the results. We study four models: 
critical percolation ${\cal LM}(2,3)$ and the logarithmic Ising model ${\cal LM}(3,4)$
but also two which do not have Potts model cousins, namely the model 
${\cal LM}(3,5)$ with $c=-\frac35$ and $\kappa=\frac{20}3$
and the logarithmic tricritical Ising model ${\cal LM}(4,5)$ with $c=\frac7{10}$ and $\kappa=5$.
The results for the latter models demonstrate that fractal dimensions can be measured also
for models not having Potts model cousins. 
For all four models, measurements are made of $d_D^{H\times V}$ 
for several lattice sizes in two different ways, as described in Section~\ref{sec:TI}. 
The results are plotted in 
Figure~\ref{fig:dimDefect}. Fits are made for both sets, discarding the smallest lattices. 
The full curves go from the smallest to the largest lattices used for the fit. The fitted curves 
are extended with dashed lines to the rejected measurements. Numerical results appear in 
Table~\ref{tab:resultsDM}. The datum $1.835|1$, for example, 
means that the $95\%$-confidence interval is of 
$1$ unit of the last digit of $1.835$. We stress that this is the interval due to statistical uncertainty 
of the measurements. The asymptotic behaviour is very good for the models
${\cal LM}(3,5)$, ${\cal LM}(2,3)$ and ${\cal LM}(3,4)$. There is a slight departure from the 
predicted theoretical value in the case of the logarithmic tricritical Ising model
${\cal LM}(4,5)$. 
One notices that the $95\%$-confidence interval does not always contain the theoretical 
prediction. The analysis in \ref{sec:extract}, however, will show how delicate the implied extrapolation is. 

\begin{table}[h!]
\begin{center}\leavevmode
\begin{tabular}{|c||c|c||c|c|}
\hline
&&&&\\
model             & $\kappa$       & $d_D^{\text{\rm theo}}=1+\frac{\kappa}8$   & $\widehat{d_{D,1}}$  & $\widehat{d_{D,2}}$  \\
&&&&\\
\hline
&&&&\\
${\cal LM}(3,5)$ & $\frac{20}3$   & $\frac{11}6\sim 1.833...$ & $1.835|1$ & $1.835|1$ \\
&&&&\\
${\cal LM}(2,3)$       & $6$            & $\frac{7}4= 1.75$      & $1.75059|62$ & $1.75133|58$ \\
&&&&\\
${\cal LM}(3,4)$      & $\frac{16}{3}$ & $\frac{5}3\sim 1.667...$  & $1.665|1$ & $1.666|1$ \\
&&&&\\
${\cal LM}(4,5)$ & $5$            & $\frac{13}8= 1.625$    & $1.618|1$ & $1.620|1$ \\
&&&&\\
\hline
\end{tabular}
\end{center}
\caption{Values $\widehat{d_{D,1}}$ and $\widehat{d_{D,2}}$ with $95\%$-confidence intervals.}
\label{tab:resultsDM}
\end{table}

The setting for our simulations is quite different from those used in previous 
measurements~\cite{AAMRH,JankeSchakel05d}. Here are the main characteristics of our setting.
\begin{itemize}
\item The linear size $R$ of the objects is determined by the lattice size $H\times V$.
\item All $N_\Delta$'s considered for the averages characterize objects spanning the whole 
lattice. These objects might not be the largest among all loops but, with high probability, will be 
among the largest. In a sense, they describe only percolating objects. 
\item Also with high probability, the objects will bounce more than once on the boundary. They
will therefore probe both the bulk and the boundary and hence carry information about the
conditions of both regions.
\end{itemize}
It is interesting to recall the main features of the other simulations~\cite{AAMRH,JankeSchakel05d}
alluded to above and contrast them with those of ours. 

Asikainen {\em et al}~\cite{AAMRH} perform measurements on a single square 
lattice of linear size $4096$. In fact, they use two sets of measurements as they probe both open 
and periodic boundary conditions. They consider all FK graphs contained in a configuration. 
For each of these graphs, they calculate both its mass (the equivalent of $N_\Delta$ above) 
and its gyration radius. The masses of all FK graphs of all configurations having their radius 
in a given window $[R_i,R_{i+1}]$ will contribute to this linear size bin. The characteristics of 
their setting are the following.
\begin{itemize}
\item The linear size $R$ depends on the FK graph being studied, not on the lattice.
\item All objects of every size in a configuration are considered.
\item Smaller FK graphs may completely avoid the boundary and probe only the bulk. In the open boundary case, larger ones are likely to explore regions close and far from the boundary.
\end{itemize}
While our approach is close in spirit to the growth process described by SLE, one can say
that the study by Asikainen {\em et al} concerns global statistical features of FK graphs.  

The simulations carried out by Janke and Schakel~\cite{JankeSchakel05d} 
are limited to the Ising model but aimed at clean measurements. 
Like us, they perform measurements on several lattices. These are all square and their linear sizes 
range from $8$ to $512$. They measure both spin and FK clusters and, for each type of cluster, 
they count the number of sites occupied by the objects. This is a natural thing to do for spin clusters, 
but perhaps a little less natural for FK clusters. 
Contrarily to the Asikainen collaboration and us, they do not tackle directly the fractal dimensions 
of the objects they study. Instead, they extract the fractal dimensions from the 
{\em percolation strength} $P_\infty$ (the fraction of sites in the largest cluster) and the average 
cluster size $\chi$. This trick allows them to obtain very clean values for the 
fractal dimensions they measure from relatively small lattices. 
In summary, their setting is characterized by the following.
\begin{itemize}
\item The properties ($P_\infty$ and $\chi$) are studied as functions of the linear size of the lattices.
\item $P_\infty$ probes one particular object, similar, yet not equal, to the defect we are 
concentrating on. The observable $\chi$ is an average over all clusters.
\item They use only periodic boundary conditions and therefore all their objects are in the bulk.
\end{itemize}

As one can see, the three experiments (Asikainen {\em et al}'s, Janke and Schakel's, and ours) are 
quite different. The fact that they nevertheless give very similar results is an indication that the physical
observables considered are rather robust, with variations in boundary conditions or even definitions 
not causing major differences between measured fractal dimensions. There is no doubt, however, 
that the speed of convergence to asymptotic behaviour obtained by Janke and Schakel is impressive. 
We note that an earlier paper, that of Fortunato \cite{Fortunato}, also reports the cluster mass for the Ising 
and the $3$-Potts models for the {\em spin clusters}. For the Ising model, Fortunato reproduces 
the predicted value ($\frac{187}{96}$ that appears in Table \ref{tab:allFractalDimensions}) 
to four significant digits. For his measurements, he uses only the largest cluster in a 
configuration; this is not exactly what we do (ours always ``percolate'' from top to 
bottom but they might not be the largest) but his setting is nevertheless the one closest to ours.

\subsection{Technical issues}
\label{sec:TI}

\subsubsection{Probability distribution of $d_D^{H\times V}$ and error on measurements}
\label{sec:technical}

The most obvious way of estimating\footnote{We use ``$\hat x$'' to denote an estimate of the
random variable $x$.}
$d_D^{H\times V}$ is
\begin{equation}
 \widehat{d_{D,1}^{H\times V}}=\frac1{|S|}\sum_\sigma 
  \frac{\log N_\Delta(D_\sigma^{H\times V})}{\log R^{H\times V}}
\end{equation}
where the sum runs over the sample $S$ with sample size $|S|$, $R^{H\times V}$ is the 
total number of quarter-circles accessible to the defect (including those on the boundary), while
$N_\Delta(D^{H\times V}_\sigma)$ is the actual number of quarter-circles visited by the defect 
in the configuration $D_\sigma^{H\times V}$. For a defect entering at one corner of a rectangle 
and exiting near the opposite one, the number $N_\Delta(D^{H\times V}_\sigma)$ is never 
zero and this definition works well. For other geometric objects to be studied in 
Section~\ref{sec:HMEPRB}, however, a naive extension of
this definition fails. For example, there are configurations $D^{H\times V}_\sigma$ whose 
number of red bonds is zero and the logarithm cannot be taken for this $D^{H\times V}_\sigma$. 
An alternative is
\begin{equation}
 \widehat{d_{D,2}^{H\times V}}=\frac{\log \frac1{|S|}\sum_\sigma 
  N_\Delta(D_\sigma^{H\times V})}{\log R^{H\times V}}.
\label{eq:dm2}
\end{equation}
Under relatively mild hypotheses on the distribution of $N_\Delta$, the two sets 
$\{d_{D,1}^{H\times V}\}$ and $\{d_{D,2}^{H\times V}\}$ should converge to the same $d_D$ 
when the mesh of the lattice goes to zero, that is, when $R^{H\times V}\rightarrow \infty$. 
The rates of convergence, on the other hand, may of course differ.

\begin{figure}[h!]
\begin{center}\leavevmode
\includegraphics[width = 0.8\textwidth]{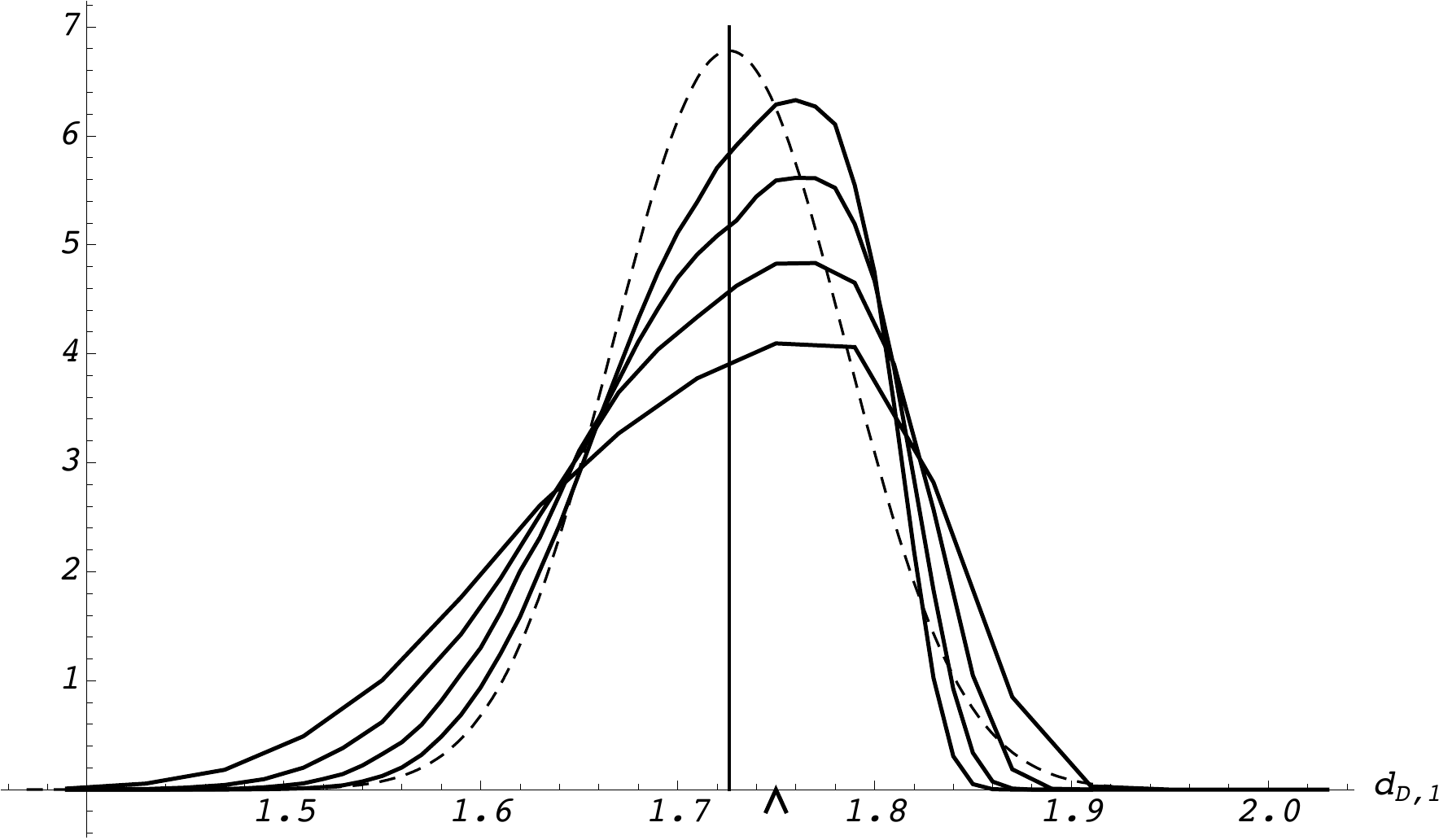}
\end{center}
\caption{Probability density functions (pdf) for $d_{D,1}^{H\times V}$ for 
$H\times V=33\times 32, 65\times 64, 129\times 128$ and $257\times 256$. 
The dashed curve represents the pdf of the normal distribution with same mean and variance 
as the distribution for $d_{D,1}^{257\times 256}$.}
\label{fig:nqDis}
\end{figure}

The probability distribution function (pdf) for $d_{D,1}^{H\times V}$ for critical percolation is shown in 
Figure~\ref{fig:nqDis} for four sizes of the lattice. The narrower distributions correspond to larger 
lattices (smaller mesh). The dashed curve is the normal distribution that has the same mean 
and variance as the distribution for $d_{D,1}^{257\times 256}$. Up to an affine transformation, 
the distribution of $\log N_\Delta$ is the same as that of $d_D^{H\times V}$. If the lattice model 
at $\beta=1$ indeed has $SLE_6$ as its scaling limit, the Hausdorff dimension should be 
$\frac74$ by Beffara's theorem (\ref{Beff}).
The figure does not exclude this, but the convergence toward the 
Dirac mass $\delta(d_D-\frac74)$ is rather slow. Even for $H\times V=257\times 256$, 
$\log N_\Delta/\log R$ is likely to take any value between $1.6$ and $1.8$. The standard deviation 
is therefore large for $\widehat{d_D^{H\times V}}$ and consequently for $\log N_\Delta$. 
Note finally that, even though the pdf for $d_D^{H\times V}$ is skewed to the right, the window 
$[\widehat{d_{D,1}^{H\times V}}-2s, \widehat{d_{D,1}^{H\times V}}+2s]$, where $s$ is the 
standard deviation of the sample, is a fair approximation of the $95\%$-confidence interval. 
To determine the confidence interval for $\widehat{d_{D,2}^{H\times V}}$, we first compute it 
for $\widehat{N_\Delta}$ and then take the $\log$ of both extremities of the interval. 
Taking the $\log$ should make the interval asymmetrical with respect to the mean. 
However, the intervals are small and this asymmetry can be ignored when 
estimating the confidence interval. 

Table~\ref{tab:confInt} contains information on the confidence intervals for critical percolation 
${\cal LM}(2,3)$ and the logarithmic tricritical Ising model ${\cal LM}(4,5)$.
As discussed in Appendix~\ref{App:upgrade}, critical
percolation is the fastest to simulate and we have large samples. 
The logarithmic tricritical Ising model, on the other hand, is the slowest that we have studied 
and samples are considerably smaller. One can see that the confidence intervals for 
$\widehat{d_{D,1}^{H\times V}}$ and $\widehat{d_{D,2}^{H\times V}}$ are essentially identical. 
Note that the precision of measurements are very good with five or six significant digits. 
For our purposes, however, they are only acceptable. Indeed, suppose again that Beffara's 
result applies to the thermodynamic limit of these lattice models. Then, around $H\times V=257\times 256$, $d_{D,1}^{H\times V}$ remains more than $0.03$ away from $d_D=\frac74$ for ${\cal LM}(2,3)$
(resp. $0.07$ away from $d_D=\frac{13}{8}$ for ${\cal LM}(4,5)$), 
but between $H\times V=257\times 256$ and $513\times 512$, this gap has been reduced 
only by $0.002$ (resp. $0.00032$). The large variance and the slow progress of the mean 
towards the asymptotic value of $d_D$ make this measurement difficult. As the authors 
of~\cite{AAMRH} point out: 
``{\em The asymptotic power law dependence of the various masses on $R$ is 
approached relatively slowly.}''

This makes the extrapolation $H,V\rightarrow\infty$ not only necessary but also delicate. 
How we perform this extrapolation is the topic of Paragraph~\ref{sec:extract}.

\begin{table}[h!]
\begin{center}\leavevmode
\begin{tabular}{|c|c||c|c|c||c|c|c||}
\hline
&&& &&& &\\
  &   &       &   ${\cal LM}(2,3)$               &                        &       &   ${\cal LM}(4,5)$         & \\
&&& &&& &\\
\hline
&&& &&& &\\
$H$ & $V$ & $|S|$ & $\widehat{d_{D,1}^{H\times V}}$ &  
 $\widehat{d_{D,2}^{H\times V}}$ & $|S|$ & $\widehat{d_{D,1}^{H\times V}}$ &  
 $\widehat{d_{D,2}^{H\times V}}$  \\
&&& &&& &\\
\hline
&&& &&& &\\
 3 & 2 & 50000000      & $1.64071|08$ & $1.69283|07$ &  100000000 & $1.52870|06$ & $ 1.58952|06$ \\
 5 & 4 & 1000000       & $1.68060|36$ & $1.71113|33$ &  100000000 & $1.58394|04$ & $ 1.61629|04$ \\
 9 & 8 & 1000000       & $1.69777|27$ & $1.71982|25$ &  50000000  & $1.60240|03$ & $ 1.62413|04$ \\
 17 & 16 & 1000000     & $1.70754|21$ & $1.72500|20$ &  10000000  & $1.60952|06$ & $ 1.62595|06$\\
 33 & 32 & 1000000     & $1.71396|18$ & $1.72856|17$ &  20000000  & $1.61291|04$ & $ 1.62614|04$\\
 65 & 64 & 1000000     & $1.71915|15$ & $1.73164|15$ &  2201000   & $1.61489|10$ & $ 1.62598|10$\\
 129 & 128 & 1000000   & $1.72312|13$ & $1.73399|12$ &  206000    & $1.61609|27$ & $ 1.62564|27$\\
 257 & 256 & 1000000   & $1.72624|12$ & $1.73588|11$ &  229700    & $1.61732|22$ & $ 1.62574|23$\\
 513 & 512 & 1000000   & $1.72867|11$ & $1.73731|10$ &  64400     & $1.61764|38$ & $ 1.62513|38$\\
 1025 & 1024 & 1000000 & $1.73074|09$ & $1.73855|09$ &            &              &              \\
 &&& &&& &\\
 \hline
\end{tabular}
\end{center}
\caption{$\widehat{d_{D,1}^{H\times V}}$ and 
$\widehat{d_{D,2}^{H\times V}}$ with confidence intervals for two of the models. 
The notation $1.64071|08$ means that the $95\%$-confidence interval is $1.64071\pm0.00008$.}
\label{tab:confInt}
\end{table}

\subsubsection{Extracting $\widehat{d_D}$ from $\{\widehat{d_{D,1}^{H\times V}}\}$ 
and $\{ \widehat{d_{D,2}^{H\times V}}\}$}
\label{sec:extract}

By definition of the fractal dimension, the number of quarter-circles $N_\Delta$ visited by the 
defect is proprotional to $R^{d_D}$. Therefore, for sufficiently large $R$, one may assume that
\begin{equation}
 \frac{\log{N_\Delta}}{\log R}\approx d_D+\frac{a}{\log R}
\label{eq:linear}
\end{equation}
where $a$ is another constant to be determined from the data. The approximation (\ref{eq:linear}) 
is linear in $(\log R)^{-1}$ and a simple linear fit should be sufficient. 
However, the curves in Figure~\ref{fig:dimDefect} clearly show that the sizes of the lattices 
we work with are too small for \eqref{eq:linear} to be a reasonable fit.

A better approximation is proposed in~\cite{AA} where a renormalisation group analysis leads to
\be
 \frac{\log N_\Delta}{\log R}\approx d_D+\frac{a}{\log R}+\frac{b}{R^\theta \log R}
\label{eq:renGroup}
\ee
where $a,b$ are constants to be fitted and $\theta$ is a known exponent depending on the model. 
It is $\frac43$ for the Ising model, for example. 
Using \eqref{eq:renGroup} adds a hypothesis on the scaling behaviour of the model, 
that of the renormalisation group. Even though we did explore the quality of the fit obtained 
with \eqref{eq:renGroup}, we use the following linear fits to get $d_D$:
\be
  \frac{\log N_\Delta}{\log R}\approx \sum_{i=0}^k a_i\frac{1}{(\log R)^i}
\label{eq:paulsFit}
\ee
where the parameter $a_0$ is to be interpreted as $\widehat{d_D}$. The number $k$ of terms in 
the fit is chosen somewhat heuristically by requiring that the $p$-value\footnote{In a linear 
regression as the present ones, the $p$-value associated with one of the coefficients in the fit 
tests the hypothesis that this coefficient is zero. Therefore, a $p$-value of $0.05$ for 
the coefficient $c$ means that one might erroneously reject the hypothesis that $c$ is zero in 
$5\%$ of the cases.} for each $a_i$ remains below $0.05$ and that $k$ be the same for all models. 
We find that $k=2$ is best for our data. 

Let us summarize some typical $p$-values obtained for fits of
$\widehat{d_D^{H\times V}}$ when the fits are based on all lattices but the two smallest ones:
$3\times 2$ and $5\times 4$. In fits with functions $1$ and $\frac1{\log R}$, the $p$-values 
of the coefficient of $1$ is always smaller than $10^{-14}$ and that of the coefficient of 
$\frac1{\log R}$ is smaller than $4\times 10^{-4}$, except in one case in which it is $\approx 0.3$. 
When three functions are used, $1, \frac1{\log R}$ and $\frac1{(\log R)^2}$, the 
$p$-values of the coefficients remain small: less than $10^{-11}$ for the coefficient of $1$ 
and less than $0.05$ for that of $\frac1{\log R}$. For that of $\frac1{(\log R)^2}$, the $p$-value 
is also very small $<0.02$ for all fits except those for the Ising model. That means that, 
for the logarithmic Ising model, the quadratic term might not be necessary. 
Indeed, in Figure~\ref{fig:dimDefect}, the curves for the logarithmic Ising model are seen to be 
exactly those where the concavity 
of the fit flips from being upward to downward implying that the coefficient of the 
quadratic term must be very close to zero. 
Finally, fits including a fourth term, $\frac1{(\log R)^3}$, 
have $p$-values larger than $0.1$ for many of their coefficients, not only for the added term.

We reject the values $\widehat{d_{D,i}^{H\times V}}$ obtained for the smallest lattices. To choose which 
ones to reject, we proceed as follows. We reject the lattices whose inclusion (or rejection) cause a 
major change in the quality of the fit. Keeping the lattice $H\times V=9\times 8$, but rejecting the 
smaller ones, seems to be optimal. An indirect 
confirmation of this choice comes from the hypothesis that Beffara's theorem is valid for our lattice 
in the case of critical percolation. We also have to choose $R$.
In the definition of the Minkowski dimension, $R$ is the linear size of the box superimposed on 
the defect. An initial choice could therefore be the square root of the total number of quarter-circles: 
$\sqrt{2H\times V}$. However, the defect has access also to the half-circles on the boundary. 
A better choice is therefore the square root of the number of quarter-circles that can be occupied 
by the defect. In the present geometry, this is 
\be
 R=\sqrt{2H\times V+4V}
\label{R4V}
\ee
and this is the definition we use. This choice should not have a significant impact if large lattices 
are considered. For the sizes considered here, though, there is an effect, as discussed below.
The values reported in Table~\ref{tab:resultsDM} are those obtained with (\ref{R4V}) and
$k=2$ for both sets $\{ d_{D,1}^{H\times V}\}$ and $\{ d_{D,2}^{H\times V}\}$ based on lattices with 
$H\times V \ge 9\times 8$ only. Using \eqref{eq:renGroup} for the 
logarithmic Ising model gives $\widehat{d_{D,1}}=1.6693$ and $\widehat{d_{D,2}}=1.6652$; the quality 
of these results is comparable to the values in the table, but no better. As we shall see shortly, 
the limitation is clearly the errors on the $\widehat{d_{D,i}^{H\times V}}$ for large $H\times V$ and the 
choice of definition for the parameter $R$. 

We have explored also the possibility of predicting, from our data, both the fractal dimension 
$d_D$ and the exponent of the correction term proposed by \eqref{eq:renGroup}. 
If the correction term is small, 
this amounts to a nonlinear fit for the parameters $a_i$ in
\be
 N_\Delta=R^{a_0}a_1\left(1+\frac{a_2}{R^{a_3}}\right).
\ee
But we have only $10$ lattices ($11$ for critical percolation) for $4$ parameters. 
The Levenberg-Marquardt method (see~\cite{numerical}, for example) does provide an excellent 
fit (in the sense that the curve is very close to the data), but the exponent $d_D$ and 
$\theta$ are totally off. Asikainen {\em et al} say that they use bins $[R_i,R_{i+1}]$ with 
$R_{i+1}=\sqrt{2}R_i$. This gives them more than twenty data points. Still, like us, they claim 
that the prediction of the correction exponents is impossible with their range of data. 
We abandoned this idea.

To evaluate the statistical error on the reported values $\widehat{d_{D,1}}$ and $\widehat{d_{D,2}}$, 
we perform the following simulation. We add noise to each datum of either set 
$\{ \widehat{d_{D,i}^{H\times V}}\}, i=1,2$, distributed normally with zero mean and variance equal to the 
standard deviation measured from the previous experiment (see Figure~\ref{fig:nqDis}). 
With this new set of ``noisy'' data, we work out the fit as described above. Repeating this experiment 
a large number of times ($1000$) gives a good idea of the deviation on $\widehat{d_{D,1}}$ 
or $\widehat{d_{D,2}}$. For the best data, i.e.~those for critical percolation, the noise coming from 
the width of the probability distribution of $N_\Delta$ amounts to an error 
($=95\%$-confidence interval) of $0.0006$ on either $\widehat{d_{D,1}}$ or $\widehat{d_{D,2}}$. 
This is the error quoted in Table~\ref{tab:resultsDM}. We stress that this is the statistical error. 
But this method of evaluating the statistical error also gives an impression of the sensitivity of 
the measurements $\widehat{d_{D,1}}$ and $\widehat{d_{D,2}}$ on a single datum. 
We have repeated this experiment by changing {\em one} datum, that on the largest lattice 
$H\times V=1025\times1024$, to the value at either extremity of its confidence interval. 
This changes $\widehat{d_{D,i}}$ by a little more than two units of its fifth significant digit. 
In other words, the normal statistical error on a single datum leads to an error about a 
fourth of that on the whole set. This justifies our claim that the errors on the data for
$d_{D,i}^{H\times V}$ significantly limit the precision on the $\widehat{d_{D,i}}$'s. 
Of course, it would be nice to probe lattices whose linear size is a few order of 
magnitude larger than the ones allowed by our slow MC algorithms.

\begin{table}[h!]
\begin{center}\leavevmode
\begin{tabular}{|c||c|c|c|c|c|c|c|}
\hline
&&&& &&&\\
lattices used & $2$ to $16$ & $4$ to $32$ & $8$ to $64$ & $16$ to $128$ & $32$ to $256$ & $64$ to $512$ & $128$ to $1024$ \\
&&&& &&&\\
\hline
&&&& &&&\\
$\widehat{d_{D,2}}$ with $R$ & $1.73484$ & $1.74220$ & $1.75013$ & $1.75433$ & $1.75113$ & $1.74988$ & $1.74999$\\
&&&& &&&\\
$\widehat{d_{D,2}}$ with $R'$ & $1.74210$ & $1.74430$ & $1.74961$ & $1.75348$ & $1.75115$ & $1.75003$ & $1.75004$\\
&&&& &&&\\
\hline
\end{tabular}
\end{center}
\caption{Measurements $\widehat{d_{D,2}}$ for critical percolation ${\cal LM}(2,3)$ obtained 
from a subset of the $d_{D,2}^{H\times V}$.}
\label{tab:movingWindow}
\end{table}

Janke and Schakel~\cite{JankeSchakel05d} report an interesting feature of their data 
for the Ising model. Their measurement of the dimension of the FK hull, the one to be compared 
with our dimension of the defect, shows the following behaviour. When they limit their analysis to 
their smaller lattices ($L=8$ to $48$), the dimension obtained is $1.665$, very close to the 
predicted $d_D^{\text{\rm Ising}}=\frac53$. But when they use
their largest one 
($L=64$ to $512$), the result seems to slip down a little (to $1.641$). This is obviously a small effect. 
It should be noted that they report results, for this mass, only with the average cluster size $\chi$. 
In Table~\ref{tab:movingWindow}, we report the values 
$\widehat{d_D}$ for critical percolation obtained from the set $\{\widehat{d_{D,2}^{H\times V}}\}$ by extrapolation 
from $4$ consecutive lattices. For example, the first fit is that with 
$H\times V=3\times2$ to  $17\times16$. 
(We admit, of course, 
that using $4$ data for a $3$-parameter fit is not ideal.) The results of the upper line use 
the definition (\ref{R4V}) of $R$ as $\sqrt{2H\times V+4V}$, the definition used throughout this analysis. 
These first move up and even overshoot the theoretical value, but then come back to extremely 
good ones. The problem reported by Janke and Schakel is not seen here. As underlined in 
Section~\ref{sec:defMeasure}, our settings are different. Our defect always crosses diagonally 
the lattice and one could guess that its gyration radius is fairly uniform. It remains to be seen how this 
could be an advantage. Note that we perform their experiment with our data for 
{\em critical percolation} as these go 
to $V=1024$. The same experiment done for our data for the logarithmic Ising model show the same 
behaviour (including the overshot and the final improvement of estimates when using larger lattices).
Table~\ref{tab:movingWindow} is completed
by results done with respect to another definition of $R$, namely 
$R'=\sqrt{2 H\times V}$. 
Although this change hardly alters the fits on large lattices, it has an important effect 
on fits with small ones. This is to be expected. It nevertheless demonstrates how sensitive 
the extrapolation is to the definition of $R$. For the fits reported in Table~\ref{tab:resultsDM}, 
the switch from $R$ to $R'$ leads to a change in $\widehat{d_D}$ within the confidence interval for 
all models but the logarithmic tricritical Ising model ${\cal LM}(4,5)$
where it is responsible for an increase twice as large as 
this interval. This indicates that, for the lattice sizes studied, there is no use trying to extract 
better extrapolations from our data, unless one can choose an $R$ on a more theoretical basis.

\subsubsection{Importance of boundary half-circles}

In~\cite{PR}, the role of boundary conditions as described by defects is studied for 
critical dense polymers ${\cal LM}(1,2)$. The half-circles that are added along the boundary of 
the domain obviously change the probability of the configurations. For example, if $\beta>1$, 
configurations with many loops are favoured and those with several (shorter) paths joining two 
points of the boundary will be common. However, for critical percolation that has $\beta=1$, one may 
think that these half-circles do not play any role. This is not so. Configurations with and without 
these half-circles will have noticeably different $d_D$. 

Consider a long cylinder whose circular 
section contains an odd number of boxes. There will be (at least) one defect going from one 
end of the cylinder to the other. 
The boundary of the cylinder is now the union of two circles. The paths reaching these boundary
circles can all be closed by boundary half-circles, all but one. Or they can be left open.
In configurations without half-circles, the loops starting and ending 
at the boundary will be as many open loops
competing for space near the boundary. 
The defect that crosses the cylinder will in this case be less dense close to the boundary. 
If, however, half-circles are added, they can be used by the defect to fill the space close to the 
boundary. Consequently, the dimension $d_D$ of the defect measured on configurations 
without half-circles will be smaller than that measured on configurations with them. 
The latter is probably independent of the geometry. The former is likely to depend on it; 
for example, $d_D$ depends on the length of the cylinder (without half-circles) and tends to 
$d_D$ (measured with half-circles) when the ratio length\!\!~/\!\!~circumference tends to infinity. 
This interpretation is consistent with the growth process that is described by SLE. 
When an SLE$_6$ interface is grown on, say, the half-plane, it is allowed to bounce on the real 
axis (the boundary). The loop gas configurations with half-circles at the boundary are those 
allowing this reflection of the defect on the boundary.

\begin{figure}[h!]
\begin{center}\leavevmode
\includegraphics[width = 0.99\textwidth]{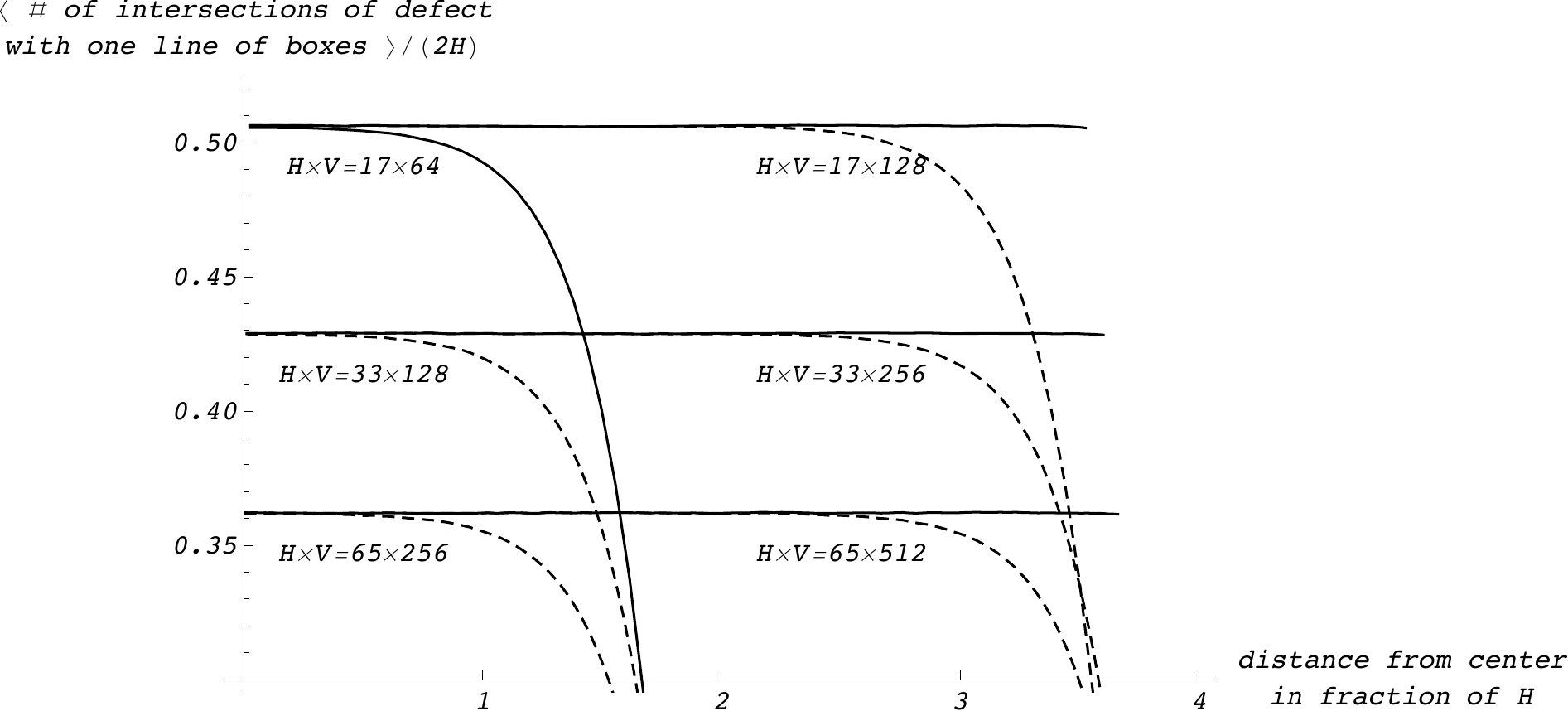}
\end{center}
\caption{Density of the defects on six cylinders as a function of the position along the cylinder axis. 
Boundary conditions are without (dashed curves) or with (full curves) half-circles.}
\label{fig:longCyl}
\end{figure}

This effect of half-circles at the boundary can be seen in simulations even on long cylinders. 
In Figure~\ref{fig:longCyl}, we plot the average of the ratio 
\be
 \frac{\text{\rm \#\ of quarter-circles visited by the defect}}{\text{\rm
total \#\ of quarter-circles}}
\ee
for each row of boxes along the cylinder. This ratio drops sharply near the boundary of the cylinder 
when half-circles are absent (dashed curves on the graph). When they are present (straight curves), 
the ratio is essentially constant on the full length of the cylinder. (The sizes given on the figure 
are labelled $H\times V$ where $H$ stands for the number of boxes along a section and $V$ for 
the length of the cylinder.) The two ratios (with or without half-circles) differ significantly over a 
distance from the extremities that is about the length of the circumference.

\section{Other geometric objects: hull, cluster mass, external perimeter and red bonds}
\label{sec:HMEPRB}

\subsection{Definition of masses for loop gas clusters}
\label{sec:otherMasses}

The classical definitions of the various masses (cluster mass, hull, external perimeter and red bonds) 
need to 
be adapted to the context of a loop gas whose thermodynamical limit for rational $\kappa$ 
is argued to be 
a logarithmic minimal model. This extension is necessary as the random variables are not spins 
but boxes with only two states. It may provide a larger domain of exploration as the loop gas  
is parametrized by the continuous parameter $\kappa$ and not only an integer, like the 
$Q$-Potts models.

\begin{figure}[!ht]
\begin{center}
\subfigure[]{\includegraphics[width=0.45\textwidth]{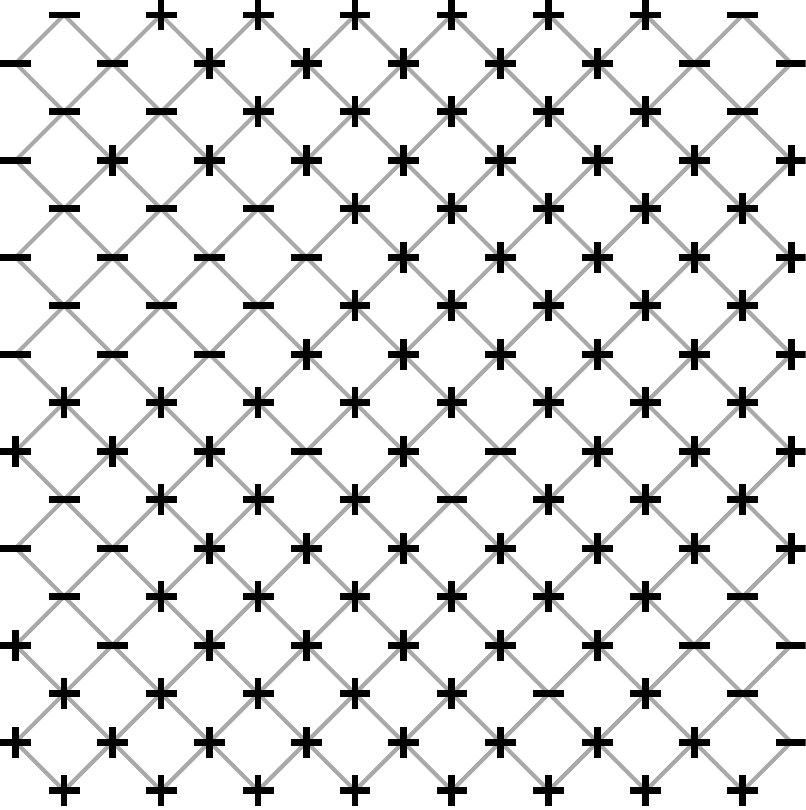}} \hfill
\subfigure[]{\includegraphics[width=0.45\textwidth]{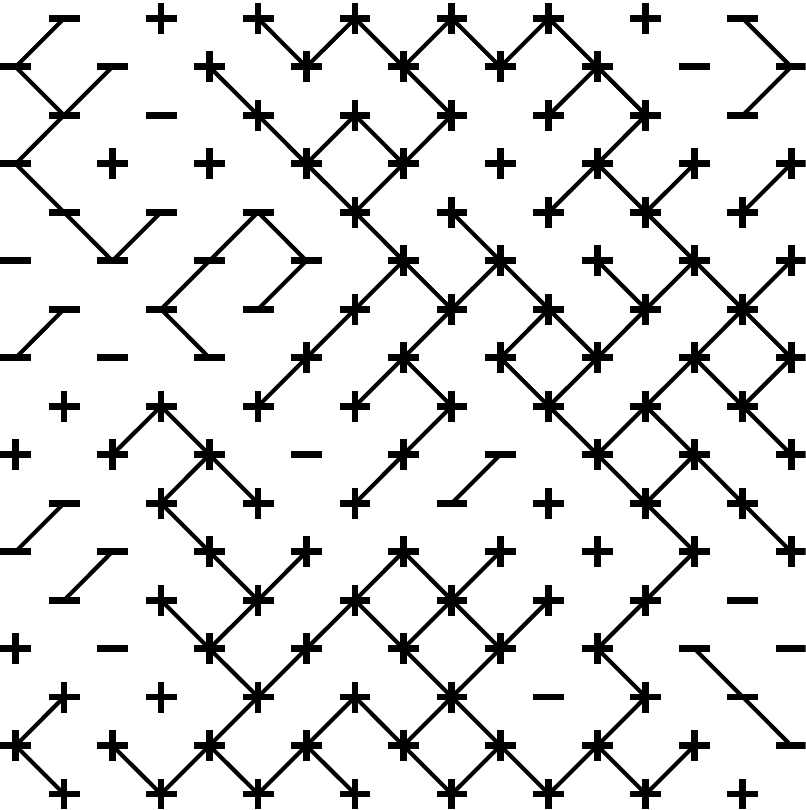}} \hfill
\subfigure[]{\includegraphics[width=0.45\textwidth]{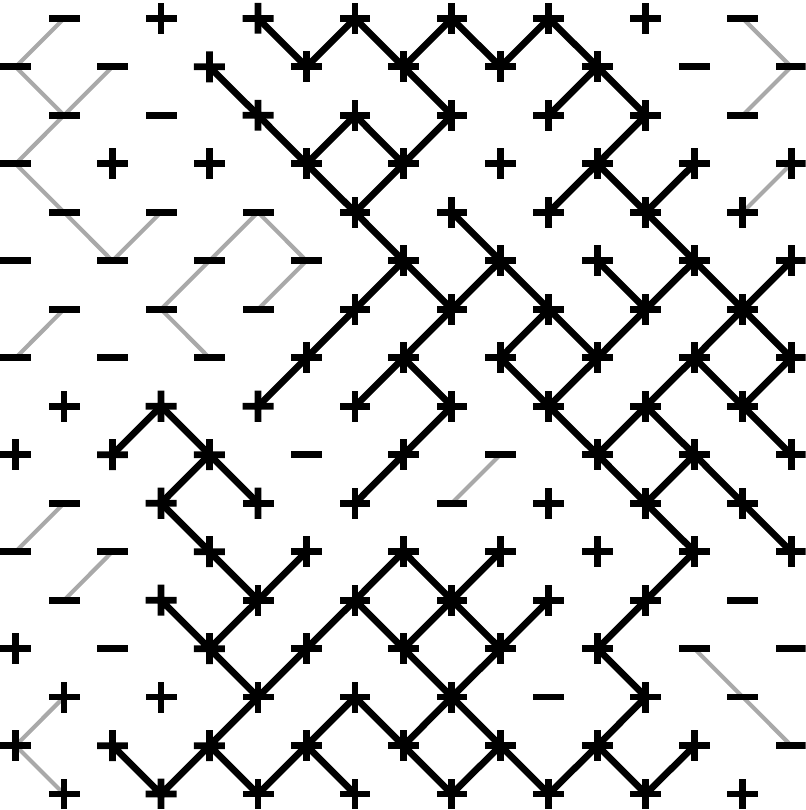}} \hfill
\subfigure[]{\includegraphics[width=0.45\textwidth]{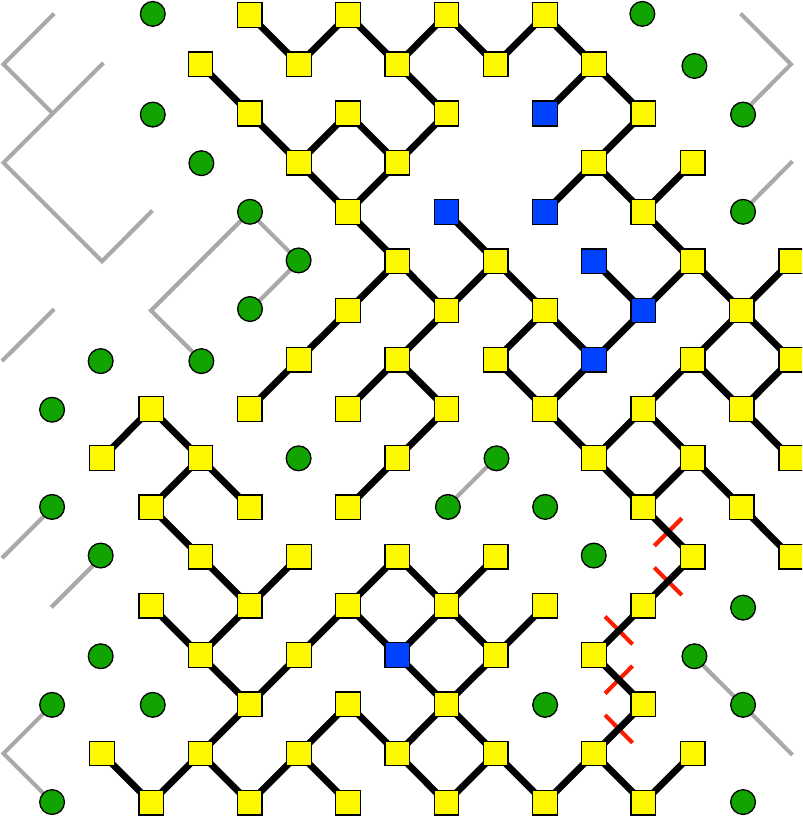}} \hfill
\caption{Four steps to construct various {\em spin masses} starting from a spin configuration.}
\label{spinMasses}
\end{center}
\end{figure}

Figure~\ref{spinMasses} (a) gives a configuration of the Ising model on a rectangular lattice. 
The reason for drawing it obliquely will become clear below. A configuration of FK clusters is 
obtained from this spin configuration as follows. Bonds between spins of distinct signs are 
erased and bonds between identical spins are removed (resp.\ kept) with probability $(1-p)$ 
(resp.\ $p$) where $p=1-e^{-\alpha}$ for $\alpha$ the inverse of the temperature 
and the Ising 
Hamiltonian of the form $H=-\sum_{\langle ij\rangle} (\delta_{\sigma_i\sigma_j}-1)$.
Figure~\ref{spinMasses} (b) shows 
a configuration obtained this way from the spin configuration 
in Figure~\ref{spinMasses} (a). 
The first step of a Swendsen-Wang upgrade consists precisely in constructing an FK 
configuration from a spin configuration. An FK configuration is partitioned uniquely into FK clusters, 
the maximal connected components of bonds and sites. We now recall the definitions of the 
various masses for these clusters. We shall concentrate on the largest cluster, 
the one shown in dark in Figure~\ref{spinMasses} (c). 
To distinguish between the classical definitions of these masses and 
the new ones to be introduced here, we add the label ``spin'' to the former and ``loop'' to the latter.

The {\em spin cluster mass} is the number of sites connected by the bonds of the FK cluster. 
A site from the cluster belongs to the {\em spin hull} if a continuous curve can be drawn from this 
site to infinity (the boundary of the lattice) without crossing any of the cluster bonds. 
The {\em spin hull} is marked by pale (yellow) squares in Figure~\ref{spinMasses} (d) and the 
other vertices of the cluster by dark (blue) squares. The {\em spin hull mass} is the number of 
sites in the spin hull, that is, the number of yellow squares. The hull partitions the whole lattice into 
three disjoint sets of sites: the hull itself, its interior and its exterior. Its interior consists of the sites of 
the lattice not belonging to the hull and from which there are no continuous paths to infinity 
avoiding all cluster bonds. The interior of the hull contains sites 
belonging to the FK cluster (drawn in dark (blue)) and others that do not. The exterior of a cluster 
is the complement of the union of the hull and its interior. The {\em spin external perimeter} (EP) 
is the subset of the exterior sites having at least one nearest neighbor in the hull. 
In Figure~\ref{spinMasses} (d), they are marked by dark circles (in green). The spin EP mass is the 
number of sites in the external perimeter. 

The last set to be described is that of {\em spin red bonds} (RB). Among the sites of the FK cluster, 
identify those that have the largest vertical coordinate and, among these, call 
$\sigma_{\text{\rm NW}}$ the site with the smallest horizontal coordinate. Similarly, define as 
$\sigma_{\text{\rm SE}}$ the site with smallest vertical coordinate and, if there is more than one
with this coordinate, the one among these with the largest horizontal coordinate. 
A bond in the FK cluster is a red bond if its removal breaks 
the FK cluster into two disconnected parts, each containing one of the two sites 
$\sigma_{\text{\rm NW}}$ and $\sigma_{\text{\rm SE}}$. In other words, if a voltage difference 
is applied between $\sigma_{\text{\rm NW}}$ and $\sigma_{\text{\rm SE}}$, the breaking of a 
single red bond will interrupt the electric current. The concept of a ``red bond'' is credited to 
Stanley. The definition appears in his paper~\cite{Stanley} but not the name. If bonds are 
imagined as fuses, the first that will turn red (and burn) are the red bonds, hence the name. 
Note that the name ``simply-connected bonds'' is also used for the red bonds. This name 
is somewhat misleading as it does not have its usual topological meaning. 
It merely refers to the fact that these bonds keep the cluster connected.
In the extension to {\em loop red bonds}, 
we shall apply the voltage difference between {\em all} topmost sites and 
{\em all} bottommost ones between the entry and exit points of the defects. 
This is a slight departure from the original definition
that we have just introduced. 
The red bonds, obtained according to this new definition, are represented on 
Figure \ref{spinMasses} (d) by small (red) slashes through them.

\begin{figure}[!ht]
\begin{center}
\subfigure[]{\includegraphics[width=0.45\textwidth]{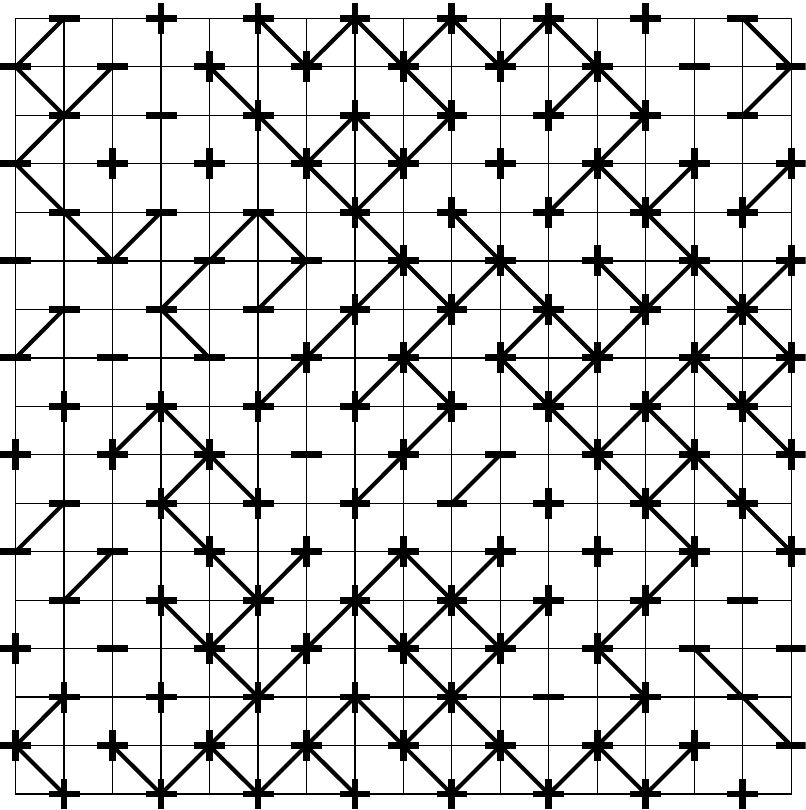}} \hfill
\subfigure[]{\includegraphics[width=0.45\textwidth]{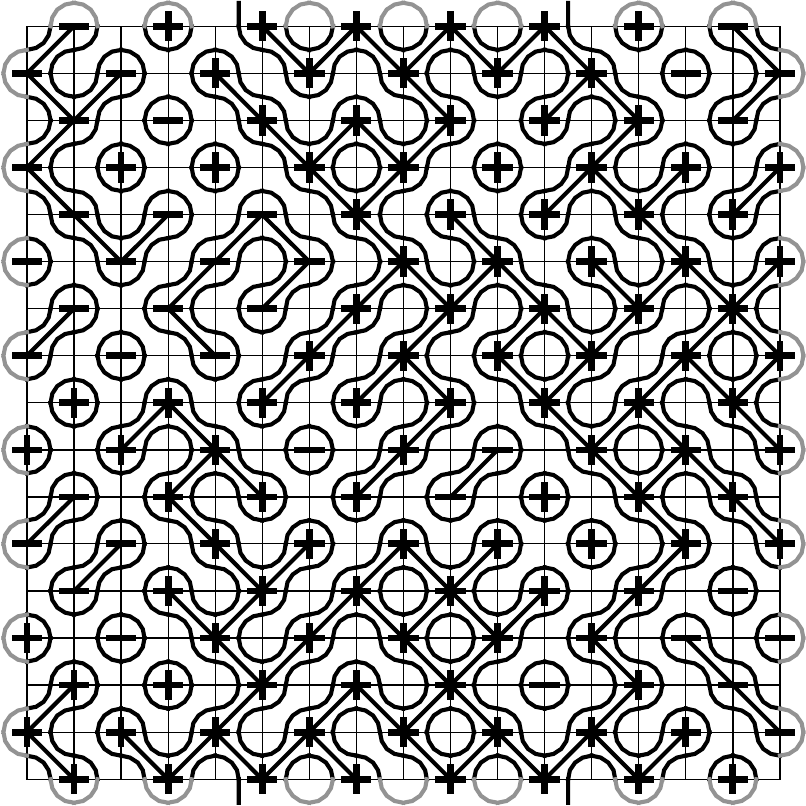}} \hfill
\subfigure[]{\includegraphics[width=0.45\textwidth]{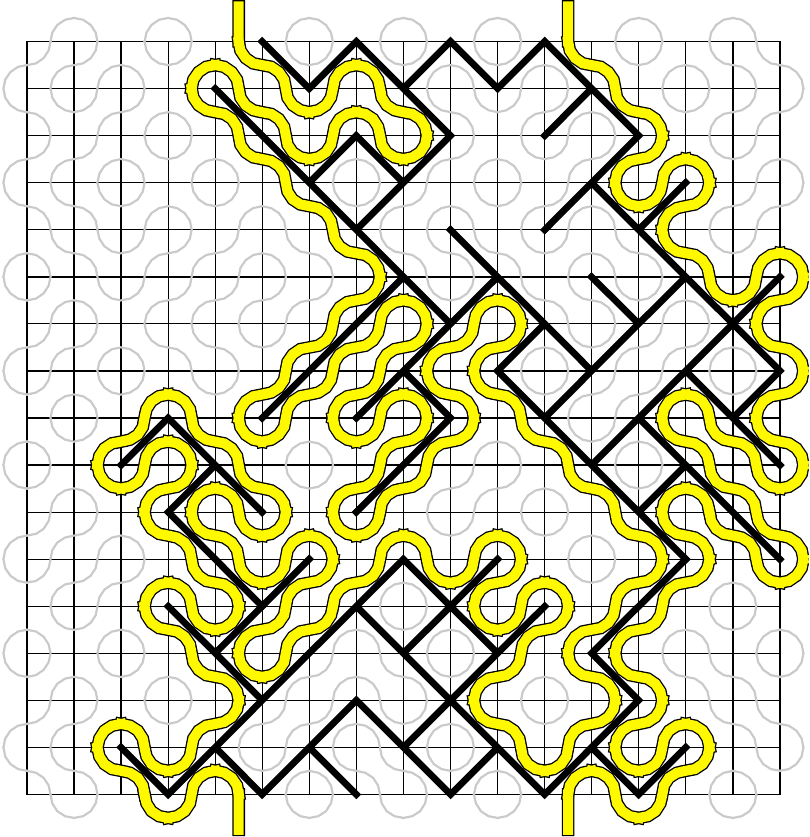}} \hfill
\subfigure[]{\includegraphics[width=0.45\textwidth]{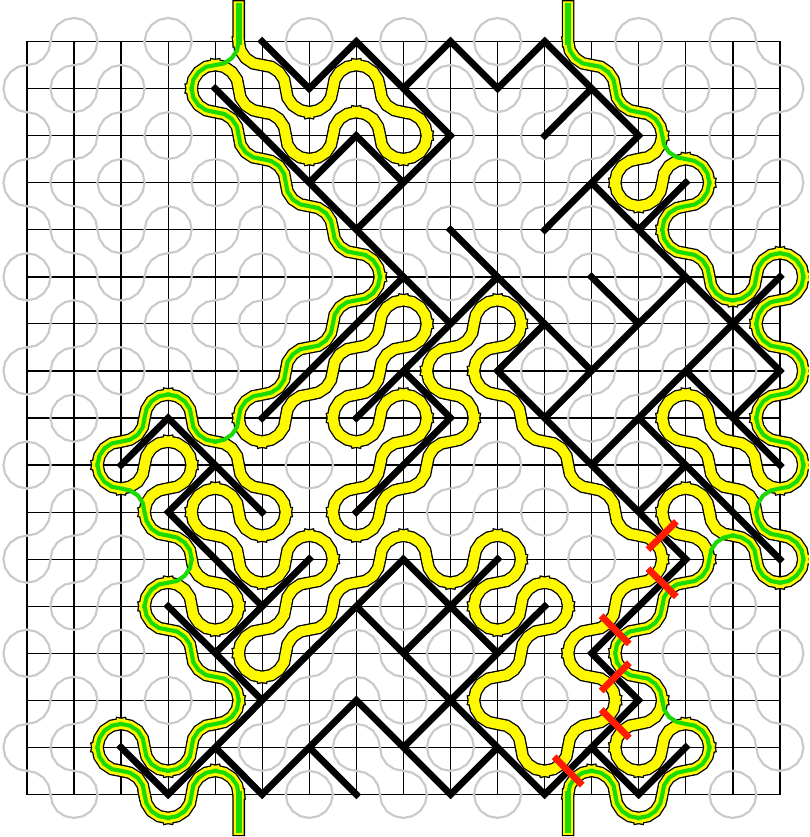}} \hfill
\caption{Four steps to construct {\em loop masses}.}
\label{loopMasses}
\end{center}
\end{figure}

The correspondence between an FK configuration (obtained from a spin configuration) and a loop 
gas configuration is one-to-one, once the boundary condition of the latter has been chosen. 
We superimpose a grid whose mesh is $1/\sqrt2$ times that of the spin lattice, as in 
Figure~\ref{loopMasses} (a). Spins of the first lattice occupy about half of the vertices of the 
new one. The state of each box in the loop gas configuration is determined from the FK 
configuration as follows. Two vertices of each box are occupied by spins. If there is a bond 
between these two sites in the FK configuration, the state of the box is such that the two 
quarter-circles do not intersect this bond. If there is no bound, then the two quarter-circles 
are drawn as if to prevent a bond to be drawn there. In such a loop configuration, the 
FK clusters not reaching the boundary are enclosed by loops as ``tightly'' as possible. 

To construct the loop gas configuration corresponding to the FK configuration of 
Figure~\ref{spinMasses} (b), we choose a boundary condition where exactly {\em two} defects
enter at the top of the lattice and exit at the bottom. 
All remaining boundary boxes are joined by half-circles between neighbouring boxes. 
The result is shown in Figure~\ref{loopMasses} (b). Again, we define the {\em loop masses} 
for the largest object (given by the defects), even though our definitions hold also for the
closed loops themselves.

The geometric objects of the loop configuration, that we now propose to replace the 
classical (spin) masses by, are based only on the loop configuration and do not entail a choice 
of a compatible spin configuration (as one would in the second and last step of a 
Swendsen-Wang upgrade). They will {\em not} agree numerically with the spin masses 
of the original finite lattice but, hopefully, will have the same thermodynamical limit.

The {\em loop hull mass} of any closed loop is the number of quarter-circles of the loop, 
including the two quarter-circles of each boundary half-circle that the loop visits. The loop 
hull of a defect is defined similarly as the number of quarter-circles from its entry to its exit. 
The loop hull created by the two defects is shown in pale (yellow) in Figure~\ref{loopMasses} (c). 

It is somewhat unfortunate that the word {\em hull} is used here as well as in the description 
of the SLE process, 
but with two different meanings. In SLE, the curve $\gamma$, which is the counterpart of a defect,
is drawn, say, on the upper half-plane $\mathbb H$.
There is no underlying lattice. 
The curve $\gamma$ is allowed to touch the real line (the boundary of $\mathbb H$) and itself.
The hull drawn by $\gamma$ is the union of the points visited by 
$\gamma$ with the set of all points that have been cut off from infinity (in $\mathbb H$) by the 
curve touching the real line or itself.
The SLE hull is therefore likely to include open domains of various sizes. 
In the loop gas, the defect never touches itself as it is the union of disjoint quarter-circles. 
The best approximation of a tangency point occurs in boxes whose two quarter-circles are 
visited by the same loop (or defect). Between the two visits of such a box, the loop has circled 
a region in the lattice whose interior could be naturally identified with an open set. 
The quarter-circles of this open set {\em do not belong} to the loop hull mass. 
This is contrasted by their counterparts in the SLE setting which {\em do} belong to the SLE hull. 

In the simulations, we will add the contributions of both defects, except for critical dense polymers 
with $\beta=0$. Counting quarter-circles of both defects amounts to visualizing these two 
defects as left and right boundaries of a percolating cluster. For polymers, configurations 
with closed loops are forbidden. This means that the defects must visit every single 
quarter-circle of the lattice. Counting quarter-circles of both defects would then give the 
total number of quarter-circles for all (allowable) configurations and the fractal dimension 
of the loop hull mass would be exactly the predicted value, $2$, and there is no need for simulations. 
Instead, we choose to restrict the counting to one of the two defects for this model ($\beta=0$). 
This raises the question of whether the fractal dimension of defects depends on the number 
of macroscopic objects in the configuration, for example the number of defects crossing 
from top to bottom. This will be discussed in Section~\ref{sec:oneOrTwo}.

\begin{figure}[!ht]
\begin{center}
\subfigure[]{\includegraphics[width=0.45\textwidth]{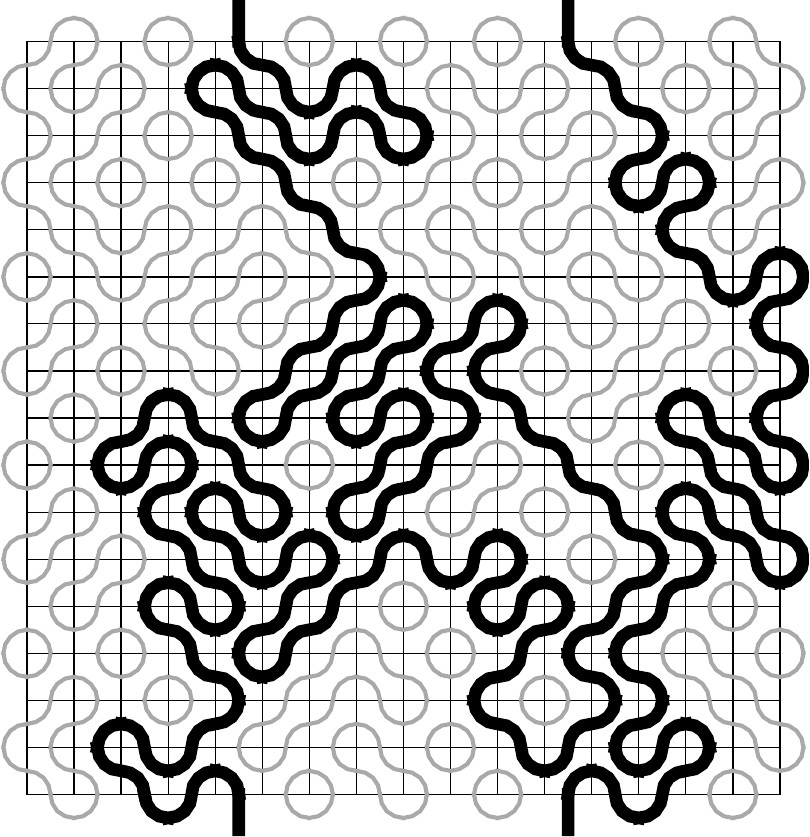}} \hfill
\subfigure[]{\includegraphics[width=0.45\textwidth]{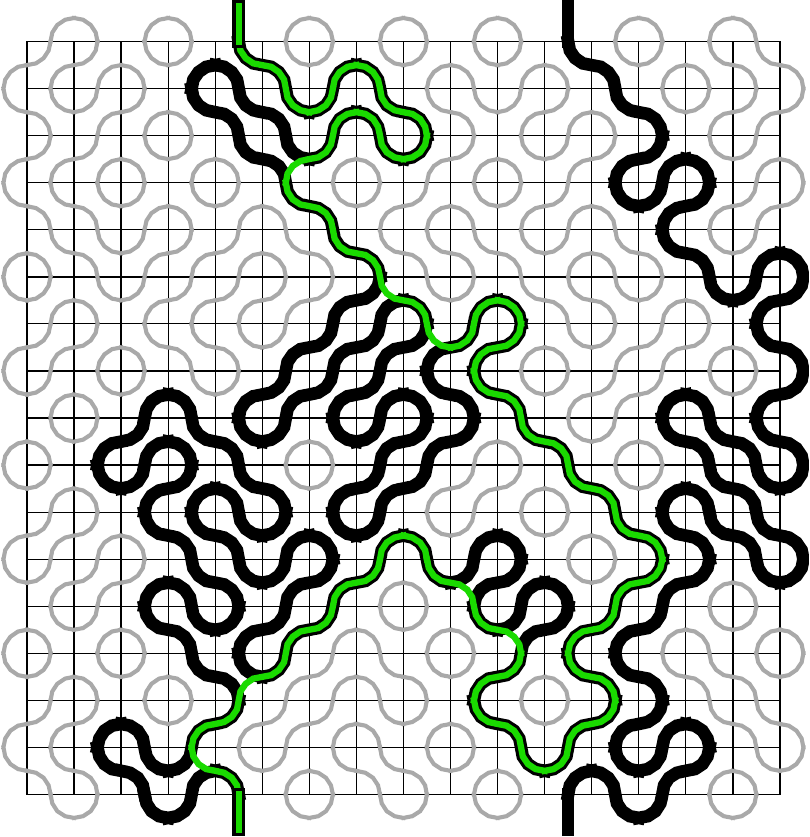}} \hfill
\subfigure[]{\includegraphics[width=0.45\textwidth]{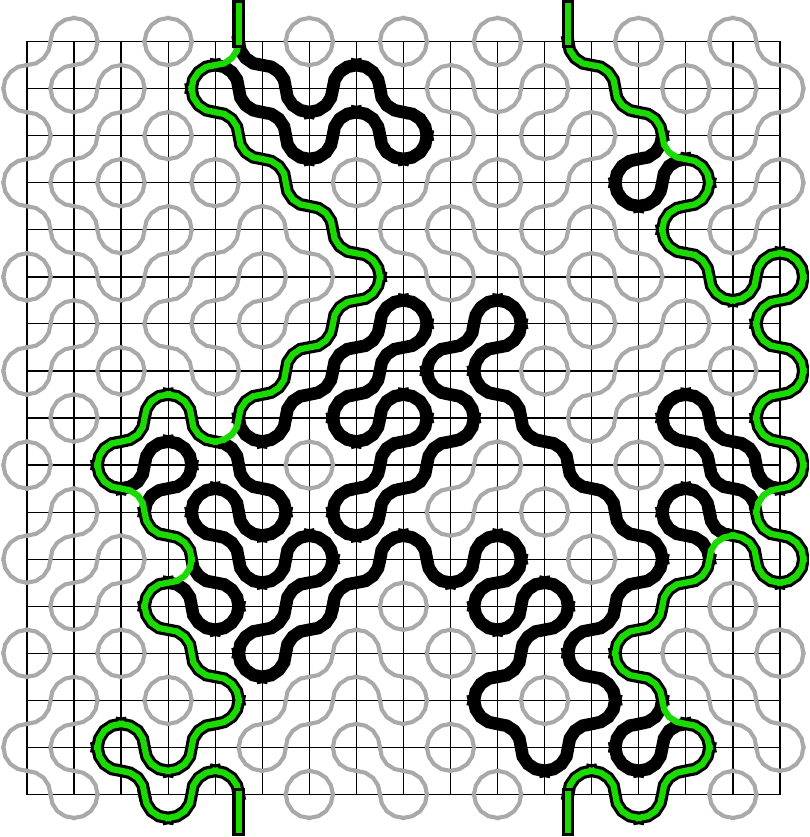}} \hfill
\subfigure[]{\includegraphics[width=0.45\textwidth]{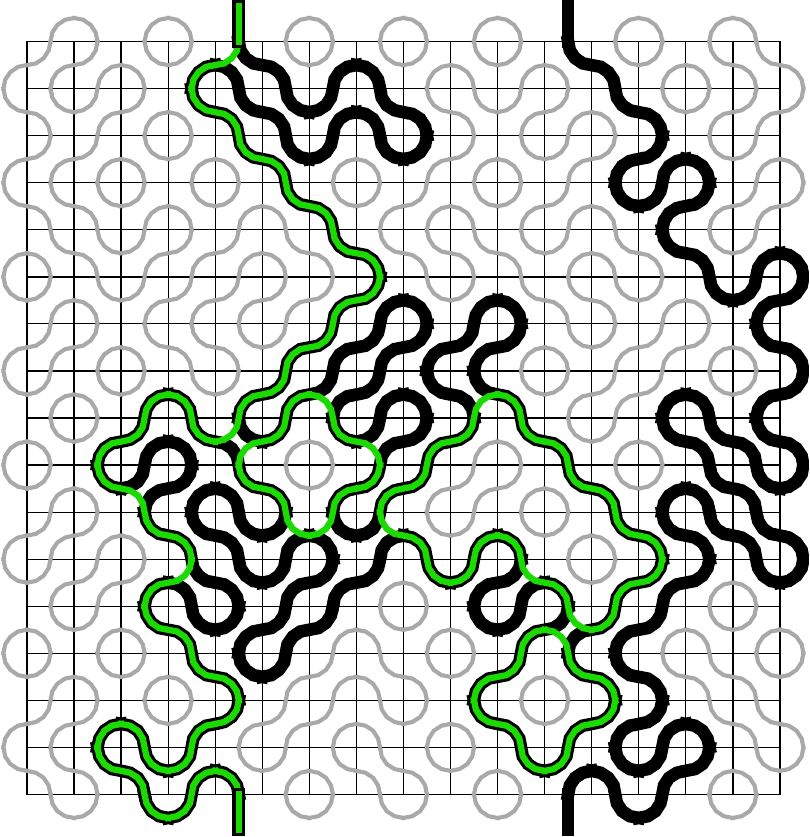}} \hfill
\caption{Three candidates for the {\em loop external perimeter}. Figure (a) shows the hull.}
\label{fig:theEP}
\end{center}
\end{figure}

The definition of the {\em loop external perimeter} (EP) is inspired by 
Grossman and Aharony's biased walker~\cite{GrossmanAharony}. When a loop 
intersects an edge, it does so perpendicularly and its direction at this intersection is 
therefore one of four choices: north, east, south and west. We start by constructing the loop 
EP of the left defect. At the top entry point of this defect, the direction is southward. 
Consider the box the loop is entering. If only one of the two quarter-circles of this box is 
visited by the loop (the defect), the external perimeter follows this quarter-circle. If {\em both} 
quarter-circles are eventually visited by the loop, the external perimeter will draw a quarter-circle 
leading to the edge on its left whether or not this quarter-circle belongs to the state of the box. 
The loop has now progressed to a new face visited by the loop. Its next step is chosen in the 
same left-biased way. The loop EP mass is defined as the number of quarter-circles visited 
by the left-biased walker following the left defect from its entry to its exit. As we are going to 
explore two other ways of defining the EP, we denote
the one just described by EP1. An example of a hull is drawn in Figure~\ref{fig:theEP} (a) 
and its external perimeter, according to definition EP1, appears in Figure~\ref{fig:theEP} (b).  
Note that the hull of the left defect is part of or to the left of its external perimeter. As can be seen 
there, the external perimeter cuts out many of the inner meanderings of the hull. To use Janke 
and Schakel's words~\cite{JankeSchakel06b}, it is a smoother version of the hull. 
Even though we have not measured the external perimeter of closed loops, we complete 
our definition EP1 for them. Find first the rightmost box edge crossed by the loop. If there is
more than one such box, any of them will do. 
This edge is always horizontal. Starting from this edge with a 
southward direction, draw the EP by following a {\em right}-biased walker. Again, the hull 
will be part of or in the interior of the external perimeter. The mass of the loop EP is again the 
number of its quarter-circles. It is smaller (or equal) to the mass of the loop hull. 

The next way of defining the external perimeter is closer to the interpretation of the two defects 
as enclosing a macroscopic FK cluster. In this interpretation both defects need to be considered 
and it is more natural to draw the EP for the left defect starting at the top and using a 
{\em right}-biased rule and then draw the EP for the right defect starting at its bottom and 
using again a {\em right}-biased rule. Figure~\ref{fig:theEP} (c) gives the result for the hull
in Figure~\ref{fig:theEP} (a). 
Note that it is also the external perimeter given as an example in Figure~\ref{loopMasses} (d) 
in mid-gray (green) superimposed on the hull in pale gray (yellow). As we shall see, this new 
definition (EP2) works equally well for all models but one. The problematic model is again 
critical dense polymers, the ``extreme model'' with $\beta=0$. As said before, the defects of this model 
fill the lattice. Using a right-biased rule to obtain the EP of the left defect would draw an EP 
wiggling along the boundary, using only the quarter-circles touching the boundary. 
Drawing the EP of the right defect starting from the bottom and using a right-biased rule 
would have the same effect. This would give the same number of quarter-circles for all 
configurations, and this number grows linearly with the (linear) size of the lattice. 
The resulting fractal dimension for polymers would be $1$. This is not the predicted value 
and this definition does not probe the fractal nature of the geometric object.

Our final proposal for defining the loop EP is denoted by EP3 and
consists in adding the ``inner bubbles''. 
The spin EP naturally adds sites that can be ``deeply enclosed'' within the cluster 
under consideration. Figure~\ref{spinMasses} (d) shows some of these bubbles. Indeed, besides 
the dark (green) disks to the left and to the right of the cluster, there is a handful of these 
disks trapped inside the cluster. We explore with EP3 the possibility of adding their 
counterparts in the loop gas case. Their addition requires several sweeps of the left- 
(or right-)biased algorithm. For example, after having drawn the external perimeter with, 
say, the right-biased definition for the left defect, we look for boxes visited by the hull 
but not by the EP just drawn. From those, we retain only boxes where one of the 
quarter-circles is not visited by the hull and where this quarter-circle is connected to 
the left lattice boundary by a path not intersecting the hull. Starting from this quarter-circle, 
we draw the right-biased EP of this ``inner bubble''. 
Figure~\ref{fig:theEP} (d) shows the 
resulting EP, with three ``inner bubbles''. As we shall see in Section~\ref{sec:loopExtPeri}, 
this last definition EP3 {\em does not} converge to the predicted value $d_{EP}$ but, 
instead, seems to converge to $d_H$.

A loop gas configuration is a choice of state for each box of the lattice and these states draw loops, 
not clusters. To define clusters from these loops and subsequently a {\em loop cluster mass}, 
we first define an {\em open bond}. 
The original spin lattice has spins on only (about) half of the vertices of 
the loop lattice. We say that a box of a loop configuration has (or draws) an open bond if its 
quarter-circles {\em do not} intersect the segment joining the two vertices of the original lattice. 
Maximally connected sets of bonds are clusters and the {\em loop cluster mass} of such a 
component is the number of open bonds. This is similar to clusters in bond percolation from 
which the name ``open bond'' is borrowed. When the boundary condition allows for defects, 
we define the cluster between two (vertical) defects as the connected set of bonds between 
the two defects reaching both the top and bottom boundaries. This set is unique unless 
there is a closed loop located between the two defects and connecting both the top and bottom of the 
lattice. In the unlikely occurence of such a loop, we would use the leftmost connected set of bonds.
The cluster mass is the number of its bonds. The geometric object used to define the loop cluster 
mass coincides with that of the spin cluster mass. However, the loop cluster mass counts 
bonds instead of vertices.

Cutting a bond in the cluster inside a hull (as just constructed) amounts to flipping the state of 
the corresponding box. If one wants to break the cluster between two defects into two connected 
parts in such a way that the electric current from the top to the bottom is cut 
(as does the cutting of red bonds), this flipping of the box will have to change the two vertical 
defects into two defects, the first connecting the two entries at the top boundary, the second 
the two exits at the bottom boundary. This occurs if and only if one of the quarter-circles of the box 
being flipped belongs to the left defect and the other to the right one. We therefore define the 
{\em loop red bond mass} (or RB mass) as the number of boxes visited by both defects. 
As will be seen from Figures~\ref{spinMasses} (d) and Figure~\ref{loopMasses} (d),
the two definitions used to draw the spin and loop red bonds almost coincide, 
the difference being of one extra loop red bond in this particular example.

\subsection{Measurements}
\label{sec:masses}

The first thing to choose is the appropriate definition of $R$. This linear size is related to the size 
$\epsilon$ of boxes in the Minkowski definition \eqref{eq:minkowski} and appears explicitly in 
our approximation \eqref{eq:minkowski2}. It counts how many geometric building blocks 
(quarter-circles, bonds, sites) the mass under consideration may occupy. The natural thing to do is 
therefore to use two distinct definitions of $R$, one for the hull and the external perimeter
\be
 R_{1}=\sqrt{2H\times V+2H+2V-4}
\ee
counting the number of quarter-circles, and one for the cluster and red bonds 
\be
 R_{2}=\sqrt{H\times V}
\ee
counting bonds in the lattice.

\begin{table}[h!]
\begin{center}\leavevmode
\begin{tabular}{|c|c||c|c||c|c||c|c||c|c|}
\hline
&&& &&& &&&\\
model & $\bar\kappa$ & 
$d_H^{\text{\rm theo}}$ & $\widehat d_H$ & 
$d_{EP}^{\text{\rm theo}}$ & $\widehat d_{EP}$ & 
$d_C^{\text{\rm theo}}$ & $\widehat d_C$ & 
$d_{RB}^{\text{\rm theo}}$ & $\widehat d_{RB}$  \\
&&& &&& &&&\\
\hline
&&& &&& &&&\\
${\cal LM}(1,2)$ & $2$ & 
                      $2$ & $2.000$ &
$\frac54=          1.250$ & $1.250$ &
                      $2$ & $2.062$ & 
$\frac54=          1.250$ & $1.258$\\
&&& &&& &&&\\

${\cal LM}(3,5)$ & $\frac53$ & 
$\frac{11}6\sim    1.833$ & $1.833$ &
$\frac{13}{10}=    1.300$ & $1.280$ &
$\frac{77}{40}=    1.925$ & $1.946$ & 
$\frac{14}{15}\sim 0.933$ & $0.956$\\
&&& &&& &&&\\

${\cal LM}(2,3)$ & $\frac32$ & 
$\frac{7}4=        1.750$ & $1.750$ &
$\frac43      \sim 1.333$ & $1.326$ &
$\frac{91}{48}\sim 1.896$ & $1.912$ & 
$\frac34=          0.750$ & $0.768$\\
&&& &&& &&&\\

${\cal LM}(3,4)$ & $\frac43$ & 
$\frac53\sim       1.667$ & $1.667$ &
$\frac{11}{8}=     1.375$ & $1.373$ &
$\frac{15}{8}=     1.875$ & $1.892$ & 
$\frac{13}{24}\sim 0.542$ & $0.557$ \\
&&& &&& &&&\\

${\cal LM}(4,5)$ & $\frac54$ & 
$\frac{3}8=          1.625$ & $1.627$ &
$\frac75=            1.400$ & $1.380$ & 
$\frac{229}{160}\sim 1.869$ & $1.884$ & 
$\frac{17}{40}=      0.425$ & $0.432$\\
&&& &&& &&&\\
\hline
\end{tabular}
\end{center}
\caption{Fractal dimensions of $d_H$ (hull), $d_{EP}$ (external perimeter), 
$d_C$ (cluster) and $d_{SC}$ (simply-connected bonds). 
For critical percolation ${\cal LM}(2,3)$, better estimates are available in
Section~\ref{sec:massesTechnical}.}
\label{tab:allMasses}
\end{table}

Figure~\ref{loopMasses} describes the setting of our simulations. Two defects enter from the 
top and exit from the bottom. Square lattices $H\times V$ are used with $H$ a power of $2$. 
For all models, the smallest lattice has $H=8$. The largest lattice for critical dense polymers 
has $H=128$, for critical percolation $H=1024$, while for all other models, the largest lattice 
has $H=256$. The left defect 
enters in and exits from column number $H/4+1$, while the right defect enters in and
exits from column number $3H/4$. This choice leaves a quarter 
of the boxes to the left of the left entry point, and a quarter to the right of the right entry point. 
The algorithm used ensures that the defect entering at the top left will exit at the top right 
or at the bottom left. The entry points account for $4$ edges of the lattice boundary; 
all other edges are matched pairwise by half-circles with a nearest neighbour edge. 
We limit our samples to configurations with defects crossing from top to bottom, like 
the configuration shown in Figure~\ref{loopMasses}. 
The size of these samples are of at least $2\times 10^5$ for each 
lattice, usually much larger for smaller lattices. For all models but critical dense polymers, 
we use the definition EP2 of the external perimeter. For critical dense polymers, we use EP1.

\begin{figure}[h!]
\begin{center}\leavevmode
\includegraphics[width = 0.99\textwidth]{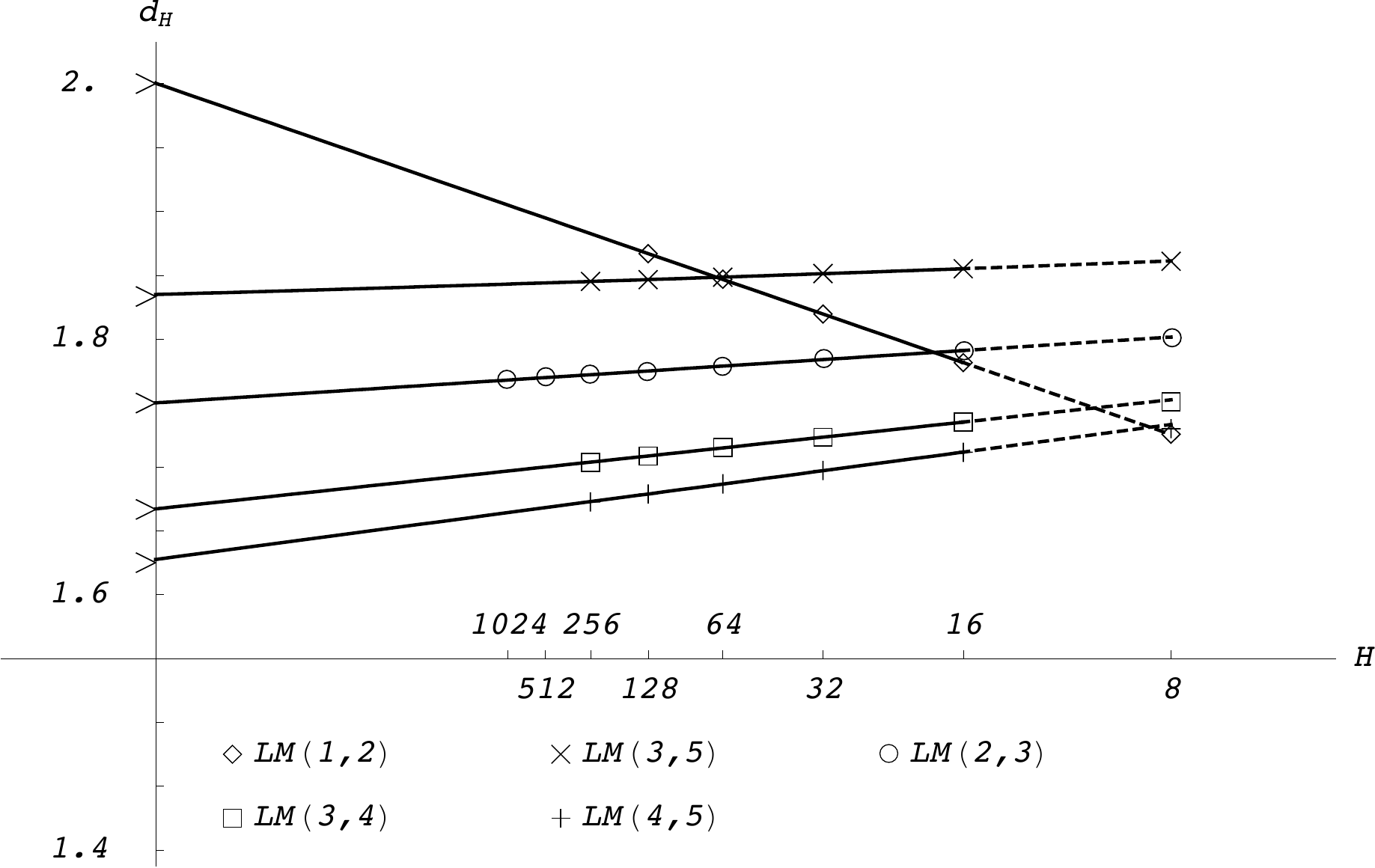}
\end{center}
\caption{Measurements of $d_H$ for the five models.}
\label{fig:fitHull}
\end{figure}

\begin{figure}[h!]
\begin{center}\leavevmode
\includegraphics[width = 0.99\textwidth]{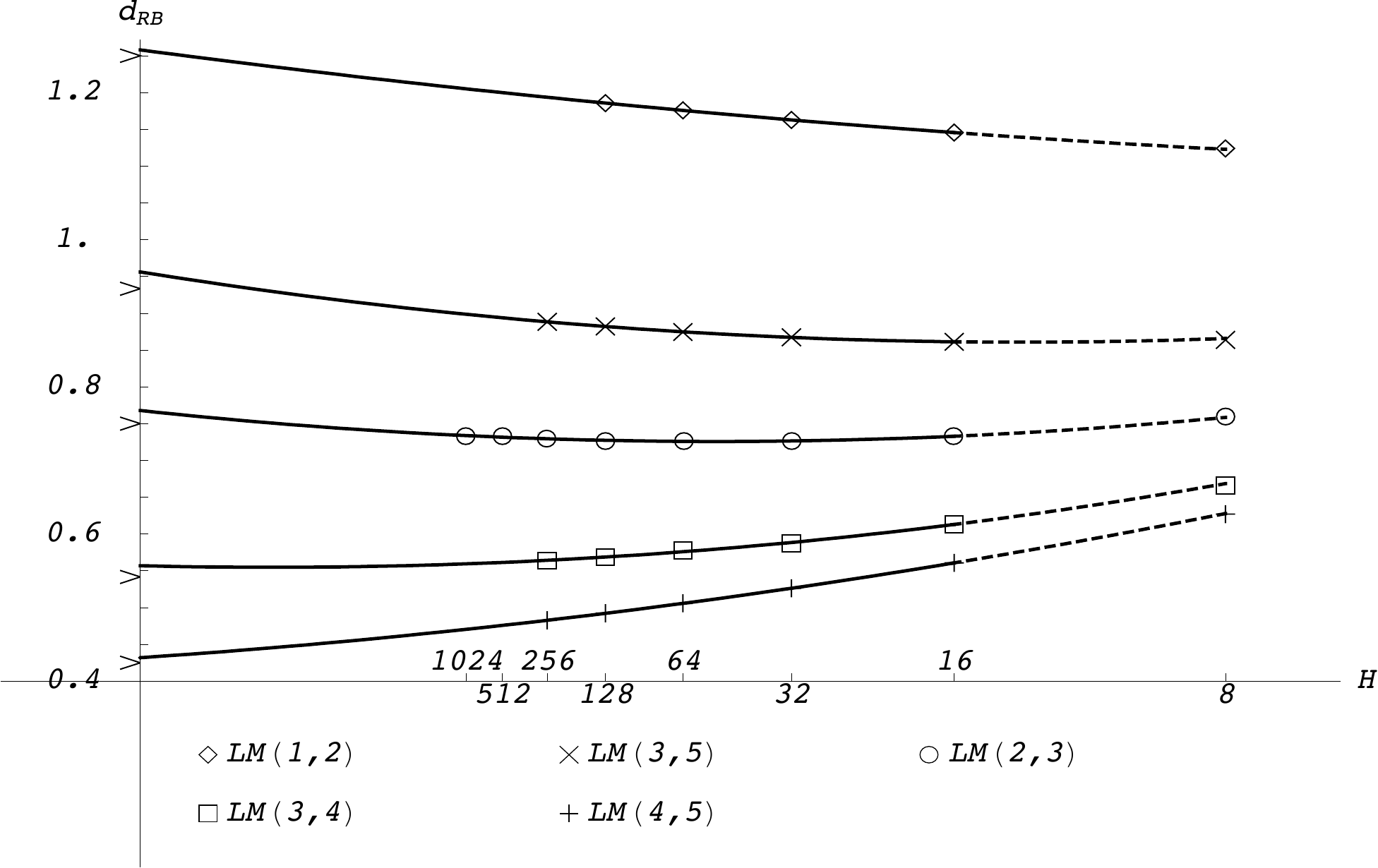}
\end{center}
\caption{Measurements of $d_{RB}$ for the five models.}
\label{fig:fitRB}
\end{figure}

The extrapolation is done as follows. For the hull mass, we do a fit with only $1$ and 
$1/\log R$ as functions and using all lattices but the smallest $8\times 8$. 
Figure~\ref{fig:fitHull} shows that the two functions $1$ and $1/\log R$ are indeed 
sufficient. The fact that the data for critical dense polymers follow a slope different from 
the slopes of the other models can be easily explained.
Recall that, for critical dense polymers, the two defects together fill the whole lattice and that, 
consequently, we choose to count only the quarter-circles of the left defect for this model. 
If we would do the same thing for one of the other models, the average number 
of quarter-circles would, by symmetry, be approximately half of what we actually measure. Therefore, 
by using formula \eqref{eq:dm2}, we would obtain 
\be
 d_H^{\text{\rm new}}=\frac{\log \langle N_\Delta\rangle/2}{\log R}=d_H-\frac{\log{2}}{\log R}
\ee
as the new measurement of $d_H$.
It turns out that, for all other models, the slope of the fit is smaller in absolute value than 
$\log 2$. This definition $d_H^{\text{\rm new}}$ would therefore lead to fits with slope 
of the same sign as that obtained for critical dense polymers. 
For the three other masses, $d_{EP}$, $d_C$ and $d_{RB}$, one observes a clear 
departure from linearity. As an example, the five curves for the red bond masses are drawn in 
Figure~\ref{fig:fitRB}. We therefore add a term $1/(\log R)^2$ in the fit. As for $d_H$, we 
reject the smallest lattice $8\times8$ and, for $d_C$, also the second smallest 
$16\times 16$. The results of the measurements, using \eqref{eq:dm2}, 
appear in Table~\ref{tab:allMasses}. 

The definition of {\em loop hull mass} that we have proposed is obtained from the number of 
quarter-circles visited by the two defects crossing the lattice from top to bottom. 
It is thus equal to the fractal dimension $d_D$ measured in Section~\ref{sec:defMeasure} 
if the object is the union of both defects crossing from top to bottom. If we are willing to accept 
that, in the thermodynamical limit, these two defects have minimal effect on one 
another, then $d_H$ should indeed be equal to $d_D$. This is confirmed by our simulations,
and the hull is seen to have the best numerical estimate of the four masses considered here.
This point will be discussed again in Paragraph~\ref{sec:oneOrTwo}. The agreement 
is less spectacular for the three other masses: $d_{EP}$, $d_{C}$ and $d_{RB}$. For these 
three, there seems to be a systematic error: the estimates for $d_{EP}$ are always smaller 
than the predicted one, and those for $d_C$ and $d_{RB}$ always bigger. It is for $d_C$ 
that the discrepancies are most obvious, especially because the five models considered 
have their predicted values all grouped in the narrow interval $[1.86, 2]$. For example, 
$\hat d_C^{\text{\rm Ising}}$ is closer to the $d_C^{\text{\rm theo}}$ for critical percolation than 
the one for the logarithmic Ising model. In Paragraph~\ref{sec:estimatesPerco}, we argue that these 
discrepancies stem mainly from the small lattices our slow 
upgrade algorithms allow us to consider.

\subsection{Technical issues}
\label{sec:massesTechnical}

\subsubsection{Estimates of critical percolation masses from smaller data windows}
\label{sec:estimatesPerco}

As said before, we choose to reject the same lattices and use the same set of fitting functions 
for all models to obtain a given mass. But bigger lattices are accessible for critical percolation and 
better estimates for this model can therefore be obtained than those reported in 
Table~\ref{tab:allMasses}. As we have done in Paragraph~\ref{sec:extract}, 
we compute estimates $\hat d$ of the various masses from a window of only $4$ lattices. 
For the four masses $d_H$, $d_{EP}$, $d_{C}$ and $d_{RB}$, we choose to restrict the fit 
functions to $1$ and $1/\log R$. The results are contained in 
Table~\ref{tab:movingWindowAllMasses}. Only estimates using \eqref{eq:dm2} are reported.
The agreement with theoretical values is better than reported in 
Table~\ref{tab:allMasses}, significantly so for all but $d_H$. These improvements 
support the hypothesis that agreement with predicted values could be improved 
also for other models by using larger lattices and, therefore, that the proposed definitions 
for {\em loop} masses have the same thermodynamical limit as their {\em spin} counterparts. 

\begin{figure}[h!]
\begin{center}\leavevmode
\includegraphics[width = 0.99\textwidth]{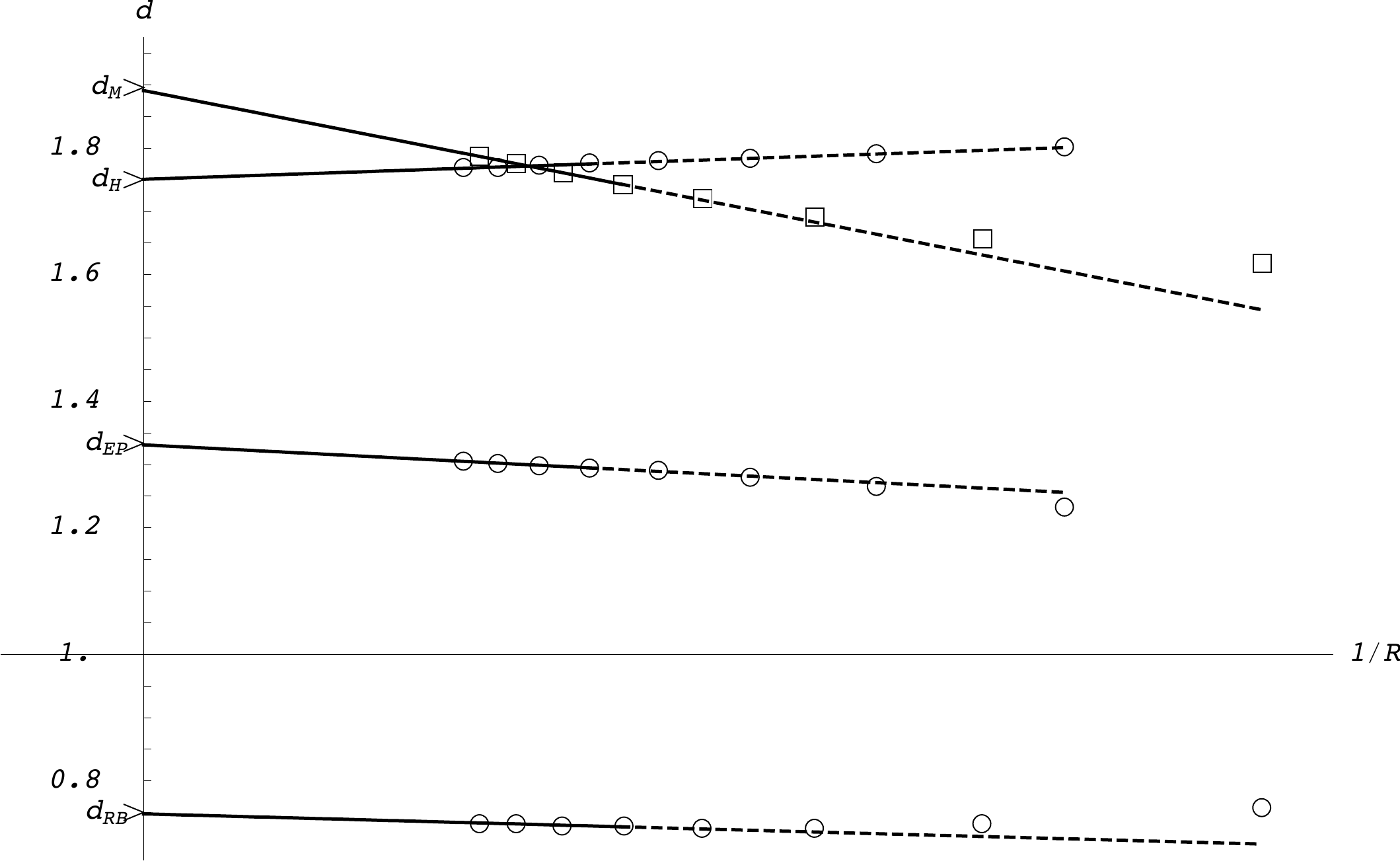}
\end{center}
\caption{Masses for critical percolation ${\cal LM}(2,3)$. 
Squares have been used exceptionnally for the cluster mass data to distinguish them from those 
of the hull. Fits are obtained using only the four largest lattices ($H=128, 256, 512$ and $1024$). 
Data for $d_H$ and $d_{EP}$ are compiled with a different definition of $R$ than those of $d_D$ 
and $d_{RB}$; this accounts for the different horizontal alignment.}
\label{fig:percoAllMasses}
\end{figure}

Figure~\ref{fig:percoAllMasses} shows the data for the four masses together with the fits 
obtained from the four largest lattices.

\begin{table}[h!]
\begin{center}\leavevmode
\begin{tabular}{|c||c||c|c|c|c|c|}
\hline
&&&& &&\\
lattices used &  $d^{\text{\rm theo}}$ & $8$ to $64$ & $16$ to $128$ & $32$ to $256$ & $64$ to $512$ & $128$ to $1024$ \\
&&&& &&\\
\hline
&&&& &&\\
$\widehat{d_{H}}$ & $\frac74=1.750$ & $1.7502$ & $1.7500$ & $1.7501$ & $1.7502$ & $1.7506$\\
&&&& &&\\
$\widehat{d_{EP}}$ & $\frac43\sim 1.333$ & $1.365$ & $1.342$ & $1.333$ & $1.330$ & $1.331$\\
&&&& &&\\
$\widehat{d_{C}}$ & $\frac{91}{48}\sim 1.896$ & $1.815$ & $1.854$ & $1.875$ & $1.886$ & $1.891$\\
&&&& &&\\
$\widehat{d_{RB}}$ & $\frac34=0.750$ & $0.6880$ & $0.7176$ & $0.7342$ & $0.7428$ & $0.7477$ \\
&&&& &&\\
\hline
\end{tabular}
\end{center}
\caption{Measurements $\hat{d}$ for the four 
masses of critical percolation ${\cal LM}(2,3)$ obtained from a subset of the data.}
\label{tab:movingWindowAllMasses}
\end{table}

\subsubsection{The definitions of the loop external perimeter}\label{sec:loopExtPeri}

In Section~\ref{sec:otherMasses}, we proposed three definitions of the {\em loop external perimeter} 
that appeared to be natural extensions of its spin counterpart. The first, EP1, counts the number 
of quarter-circles visited by a left-biased walker starting its journey at the top of the left defect and 
ending when this defect exits the lattice. The second, EP2, uses the quarter-circles visited by a 
right-biased walker traveling both defects, the left one from top to bottom and the right one from 
bottom to top. Finally, EP3 adds to the quarter-circles of EP1 the ``inner bubbles", cf. the discussion
above.

\begin{figure}[h!]
\begin{center}\leavevmode
\includegraphics[width = 0.99\textwidth]{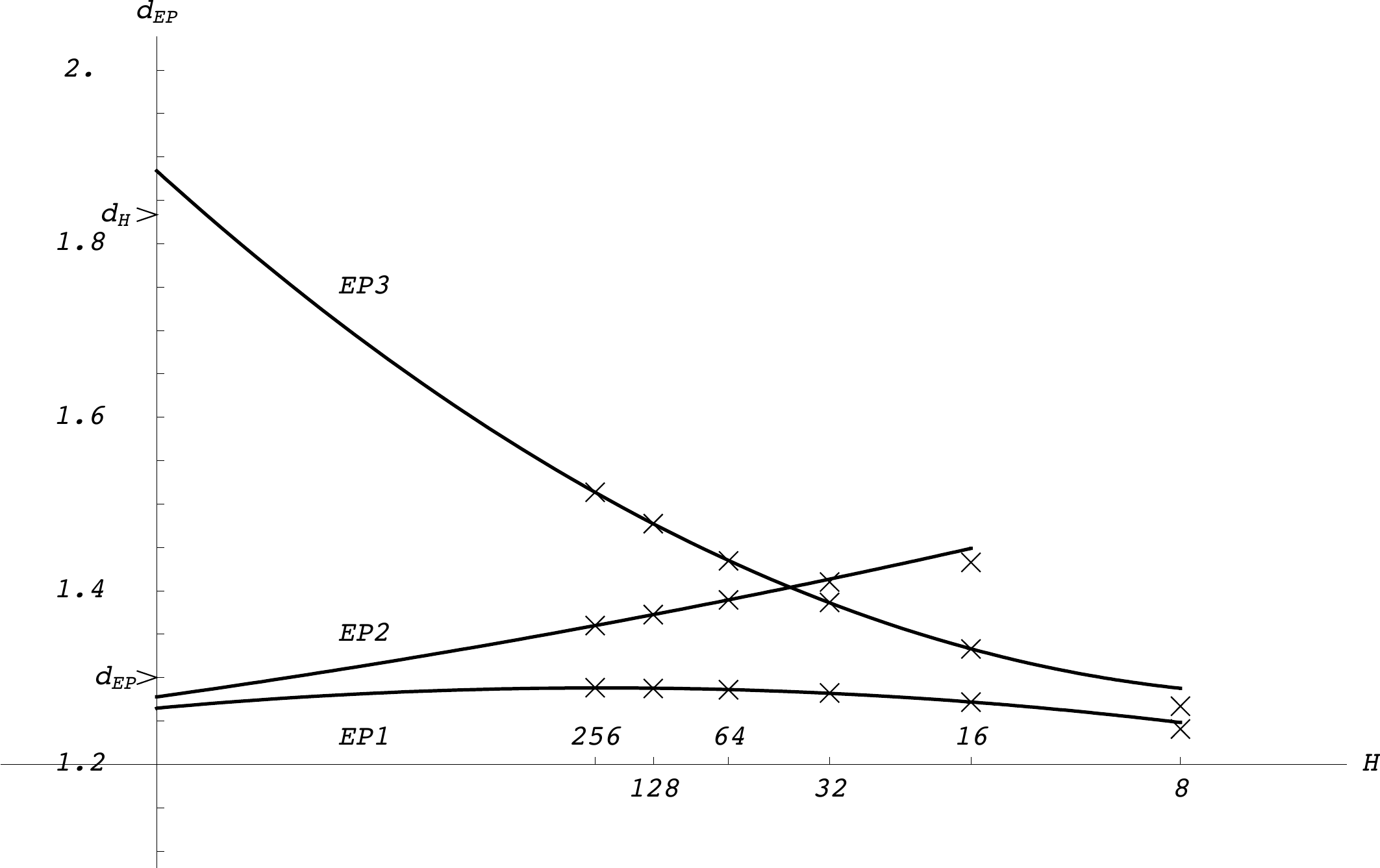}
\caption{The three definitions of the {\em loop external perimeter} for the model ${\cal LM}(3,5)$.}
\label{fig:threeEP}
\end{center}
\end{figure}

In our simulations, the definitions EP1 and EP2 appear to have the same thermodynamical limit and converge to the predicted value, whereas definition EP3 seems to behave in the limit as the hull itself. 
Even though critical dense polymers has the densest hull among the models considered,
it cannot be used to explore the three definitions of EP as some of them fail for this model.
Instead, we present here data for ${\cal LM}(3,5)$. 
After critical dense polymers, it is the model with the densest hull. The other models behave similarly
to ${\cal LM}(3,5)$.
As the densest, the hull of ${\cal LM}(3,5)$ is the one most likely to feel the boundaries and 
the two defects may impact each other. Figure~\ref{fig:threeEP} shows how the 
three definitions behave for lattices from $H=8$ to $256$. Fits with a quadratic term are 
obviously necessary, at least for EP1 and EP3; we use a quadratic term in the three fits, 
rejecting the smallest lattice in the three cases (and the $H=16$ for EP2). 
The definition EP2 explores the side of the defects closest to the boundary and one might have thought 
that the hull would hit the boundary often; this would then bring its external perimeter
EP2 closer to be linear, lowering its fractal dimension towards $1$.
This is not the case. Actually, for all lattice sizes, EP2 remains larger than EP1, 
simply because it adds the contribution of both defects. If the density of EP2 is somewhat decreased 
by the boundary, our data do not reveal this.

Obviously, the thermodynamical limit of the definition EP3 is not the predicted one with fractal 
dimension $\frac{13}{10}$. For ${\cal LM}(3,5)$ and all other models,
the extrapolation of the data we have seems to bring its limit very close to $d_H$. Because of 
the definition of EP3, $d_{EP3}\le d_H$. As it nevertheless
seems to overshoot $d_H$ a little bit, a possible 
resolution is that the definition EP3 actually {\em converges} to the hull.\footnote{The fact that 
it overshoots is of no real concern considering the extreme extrapolation that had to be done! 
Note that the data $\hat d_{EP3}^{H\times V}$ are in the range $[1.25,1.52]$ and the extrapolated 
value is $1.88$. This is not for the faint-hearted ...} 
If this guess is true, then it would mean that, 
as the lattice grows, the inner bubbles account for a fixed ratio of the hull. This is surprising 
because previous experiments did not see this. Of course, an explanation could be that the 
loop and spin external perimeters are physically different objects. But what about the definitions 
EP1 and EP2 which {\em do} reproduce the dimensions of the spin external perimeter? A second 
possible explanation is the following. The external perimeter EP3 is that of one of the largest, if not 
the largest, geometric objects in the lattice. It is conceivable that inner bubbles play a 
significant role only in the largest objects. But in other experiments, fits are done on 
objects of all sizes within the lattice.
(See for example the comment at the end of Section IV A.{} of~\cite{AAMRH}.)
Could the fits of these experiments have minimized the 
contribution of the largest objects? Yet a third explanation could be that, in adding the 
inner bubbles, we give them too much weight. Indeed, by comparing Figure~\ref{spinMasses} 
(d) and Figure~\ref{fig:theEP} (d), we see that, where there is a single ``inner spin'' of the 
spin EP, we count $12$ quarter-circles in the loop EP. 
Resolving this question requires more simulations.

\subsubsection{$d_H$ and $d_{EP}$ for configurations with one and two defects}
\label{sec:oneOrTwo}

We have just seen that, for the left defect, EPs drawn by a right-bias walker (EP2) 
or by a left-biased walker (EP1) seem to have the same thermodynamical limit. 
This rules out a significant 
effect of the boundary on that geometric object. What about the influence of the right defect on the 
left one? One could argue that the right defect acts, for the left defect, as boundary on its right side 
and therefore should have no more effect on it than the boundary. This seems to be the case.

\begin{figure}[h!]
\begin{center}\leavevmode
\includegraphics[width = 0.99\textwidth]{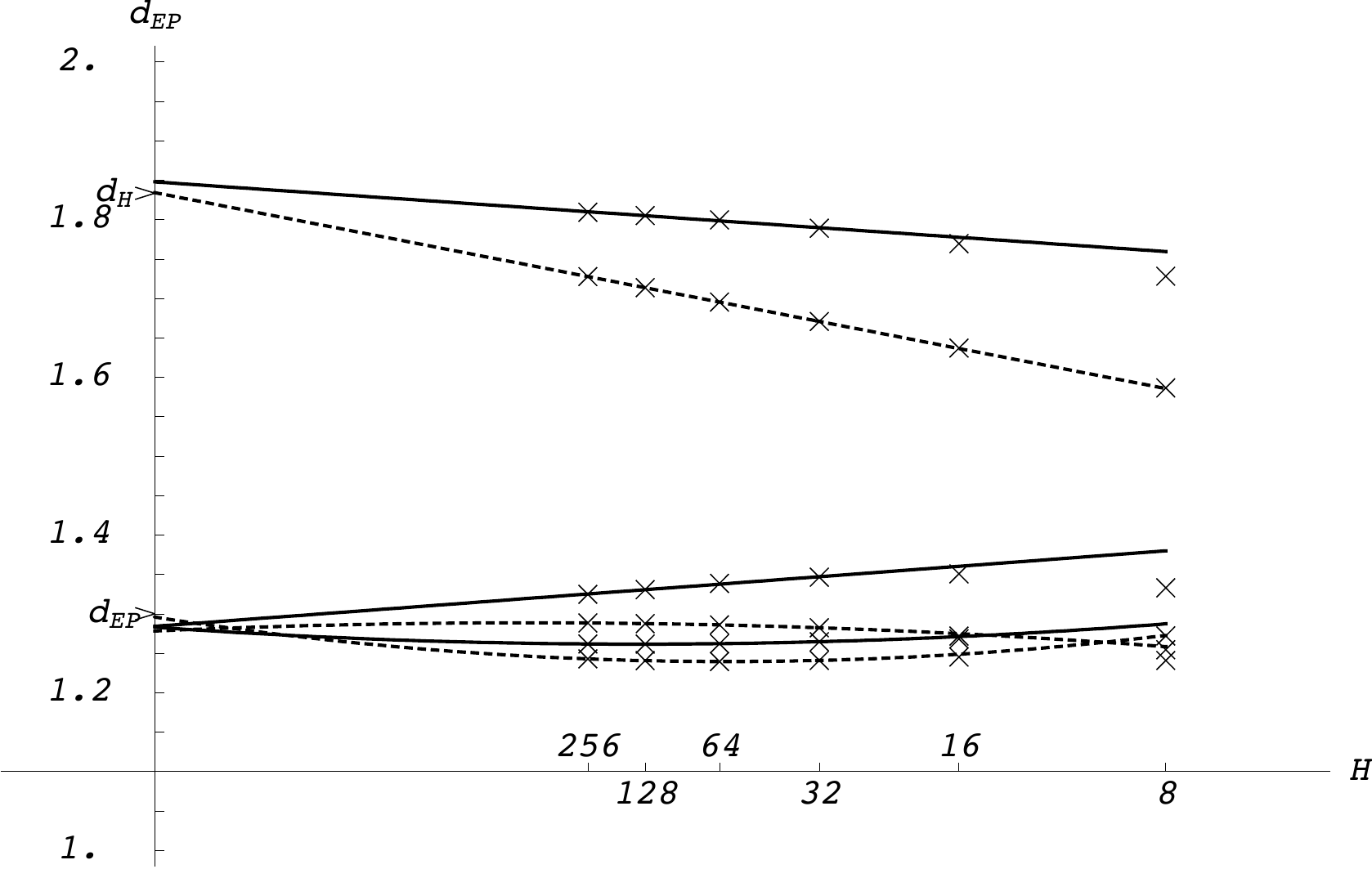}
\end{center}
\caption{Measurements for the model ${\cal LM}(3,5)$
 of $d_H$ and $d_{EP}$ of the left defect with and without the right defect.}
\label{fig:withOrWithout}
\end{figure}

Two sets of measurements are taken: one set on configurations with two defects (as we have 
done earlier) and a second one with only the left defect. The right defect is actually left in order 
not to change the size of the lattice, but is forced to remain along the top, right and bottom 
boundaries. The smallest sample is $10^5$. On these two sets of configurations, we measure 
$d_H$ of the left defect and its $d_{EP}$ with a right-biased and a left-biased rule. 
Figure~\ref{fig:withOrWithout} draws measurements of $d_H$ and $d_{EP}$ of the left defect for 
the model ${\cal LM}(3,5)$. Measurements with two defects are represented by 
dashed lines. In all cases, they lie under their counterparts measured with only one defect. 
This is natural because, in the latter configuration, the defect has more space to wander in and it 
takes advantage of this. The four curves for the external perimeter are well separated over $H=128$. 
The two obtained from a left-biased rule are over the other two with a right-biased one. 
Recall that the right-biased rule forces the external perimeter of the left defect to collide often 
with the left boundary of the lattice. Consequently, the external perimeters thus obtained tend to have 
fewer quarter-circles than the two drawn with the left-biased rule. So the order in 
which the curves appear in the figure was predictable. But what is new is that the two ways 
of measuring $d_H$ appear to have the {\em same} thermodynamical limit, and similarly for the four 
ways of measuring $d_{EP}$. This confirms our intuitive statement.

%
%

\section*{Acknowledgements} 
This work is supported by the Australian Research Council and the Canadian Natural Sciences and Engineering Research Council.

%
%

\appendix
\section{Loop gas and logarithmic minimal models}
\label{AppLoop}

\begin{figure}[h!]
\begin{center}
\ \hfill\subfigure[state $1$]{\includegraphics[width=0.075\textwidth]{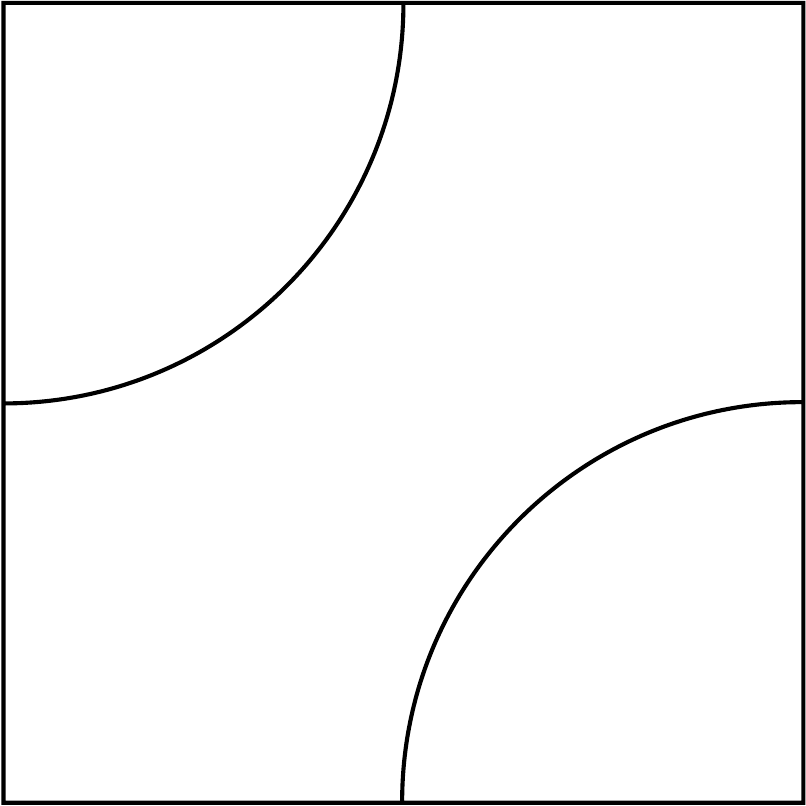}} \hfill
\subfigure[state $0$]{\includegraphics[width=0.075\textwidth]{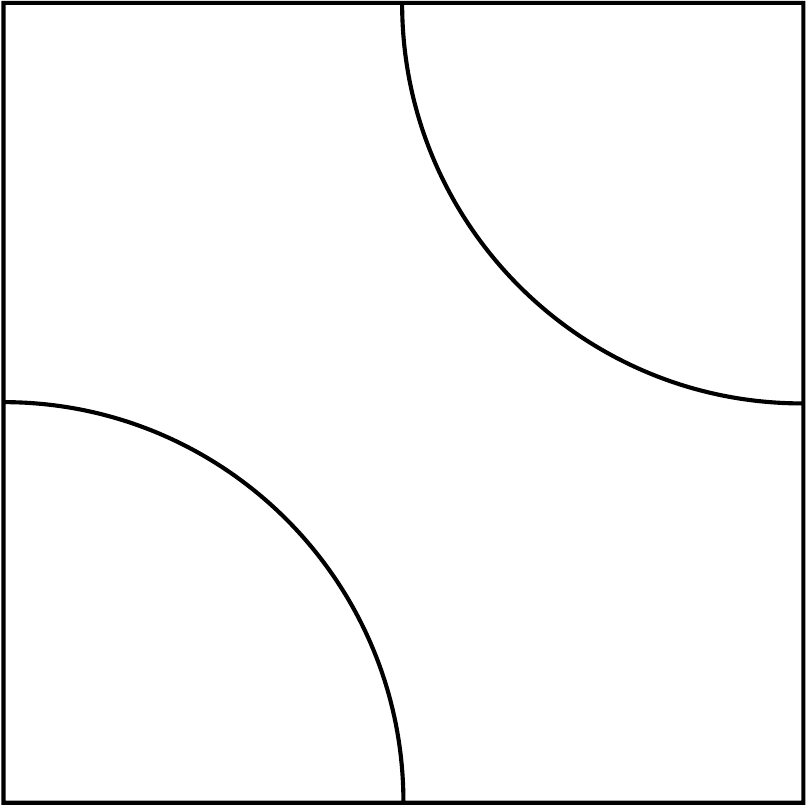}}
\hfill \
\caption{The two configurations of an elementary face with weights 
$\sigma_1=\frac{\sin(\lambda-u)}{\sin\lambda}$ and $\sigma_0=\frac{\sin u}{\sin\lambda}$.}
\label{fig:twoBoxes}
\end{center}
\end{figure}

We recall here the definitions of the loop gas and logarithmic minimal models ${\cal LM}(p,p')$ on the lattice. 
Each elementary face or box of a finite rectangular lattice, with $H$ horizontal boxes and $V$ vertical ones, is restricted to be in one of the two configurations in Figure~\ref{fig:twoBoxes} with the Boltzmann weights
\begin{equation}
\sigma_1=\frac{\sin(\lambda-u)}{\sin\lambda},\qquad \sigma_0=\frac{\sin u}{\sin\lambda}.
\end{equation}
The crossing parameter $\lambda\in(0,\frac{\pi}{2}]$ labels the given loop model with loop fugacity
\be
 \beta=2\cos\lambda.
\ee 
The parameter $u$ is the {\em spectral parameter}. For real values, it measures the spatial anisotropy of the lattice. 
The simulations in the present paper are all carried out at the isotropic point $u=\lambda/2$. 
In this case, the two box states in Figure~\ref{fig:twoBoxes}
are equally probable, as $\sigma_0=\sigma_1$.

\begin{figure}[h!]
\begin{center}\leavevmode
\includegraphics[width = 0.3\textwidth]{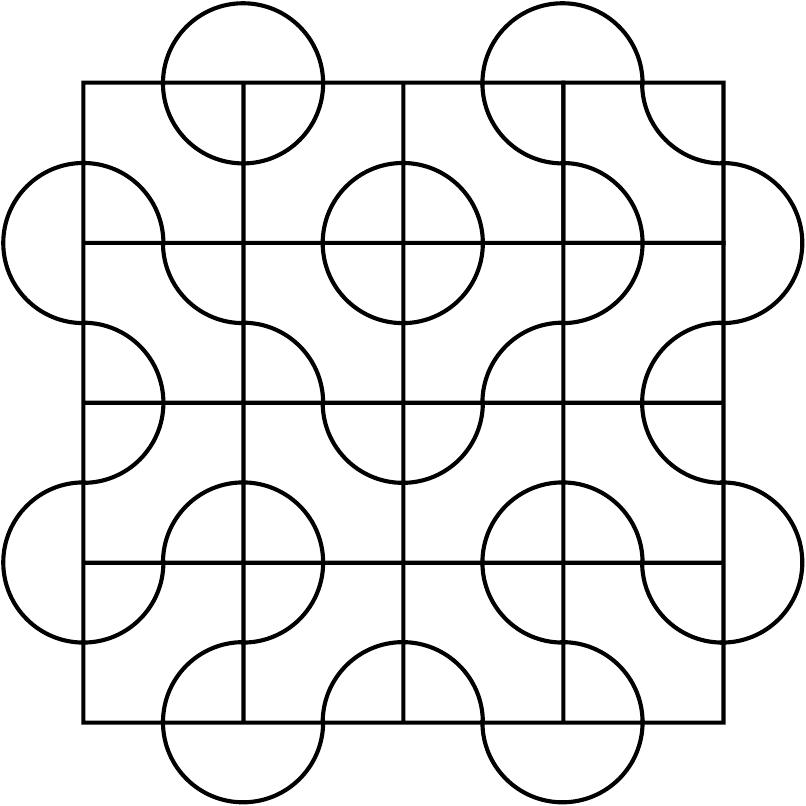}
\end{center}
\caption{A configuration on a $4\times4$ lattice.}
\label{fig:4x4}
\end{figure}

There are many ways of fixing the boundary conditions. For the
present discussion, let us join all loop ends at the four boundary segments to a 
nearest neighbour belonging to the same boundary segment. 
A $4\times4$ configuration is shown in Figure~\ref{fig:4x4}.
The Boltzmann weight of a configuration is determined by three numbers: the number $\#_1$ of 
boxes in state $1$, the number $\#_0=H\times V-\#_1$ of boxes in state $0$ and the number $\ell$ 
of closed loops. The statistical weight is then
\begin{equation}
 P(\#_0,\#_1,\ell)=\sigma_1^{\#_1}\sigma_0^{\#_0}\beta^\ell /Z
\label{eq:weightBoltzmann}
\end{equation}
where the normalization factor 
(the partition function) $Z$ assures that the sum of the $P(\#_0,\#_1,\ell)$ over all 
$2^{H\times V}$ configurations is $1$. 
In the isotropic case, the lattice model depends on the single parameter, the loop fugacity $\beta$, and 
the probability of a configuration is determined by the number of loops that it contains. 
Care has to be taken if $\beta=0$ since the partition
function itself vanishes with the boundary conditions indicated in Figure~\ref{fig:4x4}.
This does not concern us here since the boundary conditions used in the bulk part of this
paper all involve at least one so-called defect.

As a CFT, the generic loop gas is described by the Coulomb gas~\cite{Coulomb}. 
When $\frac{\lambda}{\pi}$ is rational, however, the model has some remarkable properties. Assuming $0<\lambda < \pi$, there then exist coprime integers $p$ and $p'$ such that $p'>p$ and 
\be
 \lambda=\frac{(p'-p)\pi}{p'}=(1-\frac{1}{\bar\kappa})\pi,\qquad \bar\kappa=\frac{p'}{p}.
\ee
This choice of crossing parameter defines the logarithmic minimal model ${\cal LM}(p,p')$. 
It is argued in~\cite{PRZ} that the continuum scaling limit of the loop gas 
with these special values of $\lambda$ yields a {\em logarithmic} CFT~\cite{LogCFT}, denoted by
${\cal LM}(p,p')$ and with central charge
\be
 c=1-\frac{6\lambda^2}{\pi(\pi-\lambda)}=1-\frac{6(p'-p)^2}{pp'}.
\ee
Particular examples of these continuum limits correspond to critical dense polymers ${\cal LM}(1,2)$,
critical percolation ${\cal LM}(2,3)$ and the logarithmic Ising model ${\cal LM}(3,4)$. 
It is noted that, even though this paper reports measurements only for {\em rational} 
$\lambda/\pi$, the definition of a loop gas on a finite lattice holds for {\em every} $\lambda$.

\section{Upgrade algorithms}
\label{App:upgrade}

\subsection{Critical percolation ($\beta=1$)}

Critical percolation is described by ${\cal LM}(2,3)$ where $\lambda=\frac{\pi}3$ 
and $c=0$. For the purpose of simulations, this is 
the simplest case as $\beta=2\cos\lambda=1$ and all configurations are equiprobable 
at the isotropic point where $u=\frac{\lambda}2=\frac{\pi}6$. 
At this point, we sample the configuration space according to \eqref{eq:weightBoltzmann} by 
choosing randomly the state of each box of the $H\times V$ lattice, 
giving each state in Figure~\ref{fig:twoBoxes} equal probability. 
Since $\beta=1$, no loop counting is required and obtaining good samples on large lattices 
can be done swiftly.

\subsection{Models with $0<\beta\le 2$}

For $\lambda\in[0,\frac{\pi}2)$, the loop fugacity $\beta$ ranges in $(0,2]$. 
For $\beta<1$, the probability $P$ \eqref{eq:weightBoltzmann}
can be seen as penalizing configurations with a large 
number of loops. On the contrary, for $\beta>1$, configurations with many loops are favoured. 
For $\beta\neq1$, an upgrade algorithm is needed to sample the space with the correct probability. 

We recall that the Swendsen-Wang algorithm takes advantage of 
the FK representation of the Potts model. 
For spin models, like the $Q$-Potts model, the Swendsen-Wang algorithm 
provides an efficient way to upgrade configurations. From a 
given spin configuration where each site is in one of the 
$Q$ available states, the Swendsen-Wang algorithm first constructs an FK
configuration by removing bonds between neighbouring sites that
are not in the same state and by keeping, with probability $p=p(Q)$, 
bonds between neighbouring sites in the same state. 
The FK graphs are in one-to-one correspondence
with the loop gas configurations. The second step of the algorithm
simply chooses one of the $Q$ possible states for each of the
connected components of the FK graph. But this is meaningful only
if $Q$ is an integer. The relationship between the number
of states $Q$ of the Potts model and the fugacity is
$Q=\beta^2$. Except for $\beta=1,\sqrt2,\sqrt3$ and $2$,
the corresponding $Q$ of the loop gas will not be an integer and
the second step of the Swendsen-Wang algorithm cannot be extended in an obvious
fashion.

\begin{figure}[h!]
\begin{center}
\subfigure[no entering defect]{\includegraphics[width=0.3\textwidth]{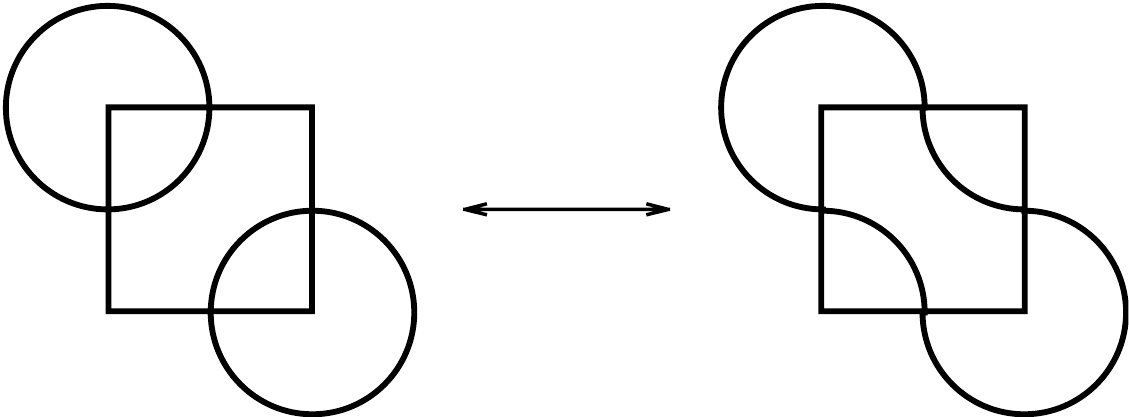}} \hfill
\subfigure[one entering defect]{\includegraphics[width=0.3\textwidth]{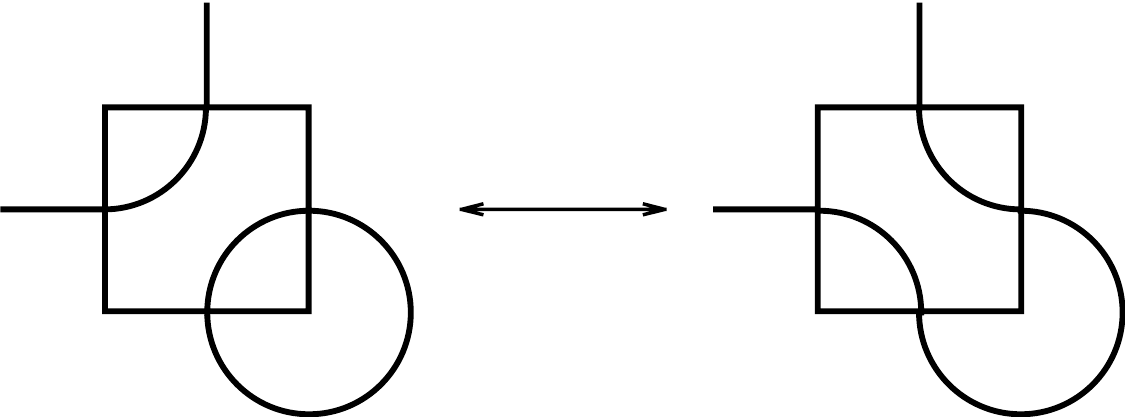}} \hfill
\subfigure[two entering defects]{\includegraphics[width=0.3\textwidth]{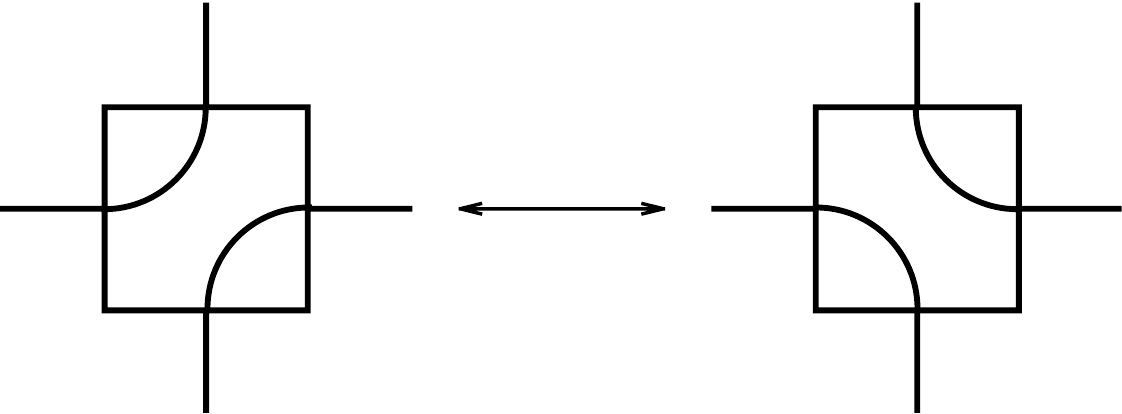}}
\caption{The effect of flipping one box on the number of loops.}
\label{fig:oneBoxFlip}
\end{center}
\end{figure}

We therefore use the traditional Metropolis algorithm, flipping
one box or a small number of them at a time. Suppose that we
have allowed defects to enter and exit the configurations. Flipping
one box will always change the number of loops by $0, +1$ or $-1$.
This is demonstrated in Figure~\ref{fig:oneBoxFlip}. The first pair
shows the situation where the two quarter-circles of the box
being flipped belong to closed loops. The portion of the closed loops
outside the box may contain a large number of quarter-circles. We represent
them as small ``ears''. The configuration before the box flip is one of the
two configurations represented in Figure~\ref{fig:oneBoxFlip} (a) and the configuration after is
necessarily the other one. The change in the number of loops is therefore
$\pm 1$. Figure~\ref{fig:oneBoxFlip} (b) depicts the case where one of the two
quarter-circles belong to a defect. The external parts of the defect are 
represented as segments. Again, the change is $\pm 1$ loop. If two
distinct defects cross the box as in Figure~\ref{fig:oneBoxFlip} (c), its flipping does not change
the total number of loops. Note that if the same defect crosses
the box twice, the situation is that of the right drawing of Figure~\ref{fig:oneBoxFlip} (b) and a 
loop will be created by the flipping process.

The simplest MC algorithm is therefore to flip one box
at a time. The box is chosen at random in the $H\times V$ lattice.
If the number of loops has not changed (as in Figure~\ref{fig:oneBoxFlip} (c)), we accept the flip. If
the number of loops has changed, the ratio 
\be
 \frac{P(\text{\rm configuration after flip})}{P(\text{\rm 
   configuration before the flip})}=\beta^{\pm1}
\label{rat}
\ee
is computed. If this ratio is larger than one, the change is accepted. If it is
smaller than one, the change is accepted with probability given by the ratio. If this algorithm
is used repeatedly, the probability of visiting any configuration starting
from any other is nonzero. This is the usual Metropolis algorithm. 

In order to calculate the change in the number of loops, it suffices to
construct the loops intersecting the box being flipped. All our simulations
involve one or two defects and these are large. 
It might therefore be useful to flip
several boxes at a time, as several could intersect defects. There is
a compromise to be struck between changing many boxes, to speed up the upgrading,
and maximizing the acceptance rate. Fortunately, the acceptance rate does not
fall too quickly as the number of boxes being flipped is increased. This
can be seen as follows. Let $x$ denote one of the $H\times V$ boxes of the
lattice and let $\sigma_x$ be a given configuration on all boxes but $x$.
The choice of state for the box $x$ gives rise to two configurations for
the whole lattice. 
If both defects cross box $x$, then the acceptance rate is $1$ since
the number of loops is unchanged. Exclude now
the case of two defects crossing $x$.
The difference between the numbers of loops between these configurations is then
$\pm 1$. If the configuration with the largest number of loops is $\sigma_x^+$
and the other $\sigma_x^-$, the ratio of their probabilities is
\be
 \frac{P(\sigma_x^+)}{P(\sigma_x^-)}=
  \frac{P(\sigma_x^+|\sigma_x)}{P(\sigma_x^-|\sigma_x)}=\beta
\ee
and the sum of the conditional probabilities is 1. We thus have
\be
 P(\sigma_x^-|\sigma_x)=\frac1{\beta+1}\qquad\text{\rm and}\qquad
  P(\sigma_x^+|\sigma_x)=\frac{\beta}{\beta+1}.
\ee
Suppose now that $\beta<1$ and then $P(\sigma_x^-|\sigma_x)>
P(\sigma_x^+|\sigma_x)$. If we start from $\sigma_x^+$, the flip of
$x$ will be accepted with probability $1$ and, if we start from
$\sigma_x^-$, it will be accepted with probability $\beta$. The acceptance rate
is therefore 
\be
 r=P(\sigma_x^+|\sigma_x)+\beta
  P(\sigma_x^-|\sigma_x)=\frac{2\beta}{\beta+1},\qquad \beta<1.
\ee
Likewise for $\beta>1$, one gets 
\be
 r=2/(1+\beta),\qquad \beta>1.
\ee 
The acceptance rate is therefore
\be
 r=2\min(P(\sigma_x^+|\sigma_x),P(\sigma_x^-|\sigma_x)).
\ee

What is the acceptance rate when several boxes are flipped at once?
Since a loop could cross more than one of the boxes flipped, the analysis
is more delicate. But if the number $n$ of boxes flipped is much smaller than 
$H\times V$, the hypothesis, that no loop crosses more than one of these
$n$ boxes, is reasonable. Let $\sigma^-$ be the configuration with the smallest number
of loops among the $2^n$ configurations under consideration. Let $p_-$ be its 
conditional probability, that is, its probability given that the other 
$(H\times V)-n$ boxes are held fixed. If $\beta<1$, then $\sigma^-$ is
the most probable and a calculation similar to the previous one leads to
the acceptance rate of
\be
  r\sim \left(\frac{1+(-1)^n}{2}\binom{n}{n/2}\beta^{n/2}
    +2\sum_{0\le i<n/2}\binom{n}{i}\beta^{n-i}\right)p_-.
\ee
The result is the same if $\beta>1$.
In either case, the probability $p_-$ is 
\be
 p_-=(1+\beta)^{-n}.
\ee 
As an example, the acceptance rate for the logarithmic Ising model ($\beta=\sqrt2$) of the flip of
a single box is $82\%$ and that of the flip of $20$ boxes is $45\%$, 
always under the assumption that no loop crosses more than one of these $20$
boxes. On large lattices ($H\times V= 128\times 128$ or larger), the measured
acceptance rate $\hat r$ is the one calculated, within measurement error. 
The assumption on loops crossing the boxes being flipped is thus confirmed as being reasonable. 
By flipping several boxes at once, we save some overhead in the upgrade process
and time when more than one of the $n$ boxes intercept the defects.

In Section~\ref{sec:masses}, we study configurations with two defects. There are two entry points 
at the top of the lattice and two at the bottom. Our study conditions the defects to enter from the top 
and exit at the bottom. The algorithm is not forcing the production of configurations with 
defects running vertically only. That is, the algorithm gives also configurations where the 
defect entering at the top left entry point exits at the top right one. We reject these and keep 
for measurements only those where the defect entering at the top left entry point exits at the 
bottom left exit point. Since upgrades are (time) expensive, we have 
explored the following way of gaining 
time. After the completion of the number of upgrades necessary for independence, we check
whether the defects are vertical. If they are not, we perform upgrades (checking between each) 
until the new configuration has vertical defects. This unfortunately biases the MC process. 
Indeed, choosing the first configuration with vertical defects after a run of configurations without them 
is likely to shift the distribution, probably onto configurations that have a constriction point 
(where the two defects have switched from a left-right to top-bottom pattern). To avoid this bias, a 
configuration without vertical defects is simply rejected and a full number 
of upgrades is redone before checking for vertical defects and possible measurement.

\subsection{Critical dense polymers ($\beta=0$)}

This case is particular because, for $\beta=0$, loops of any size are
forbidden. Defects are thus necessarily present and must fill the lattice. 
We have developed a (very slow) algorithm to generate randomly dense polymer configurations with two defects. The difficulty here is not
to sample the configuration space with the correct probability. As for 
critical percolation, all (admissible) configurations are equiprobable. 
The problem is to make sure that they do not contain loops. Suppose that we
are interested in configurations where two defects enter at the top edge 
of a rectangular lattice and exit at the bottom edge, and that a configuration
without loops is already known. The entry and exit points are fixed 
throughout the simulations. The algorithm first lists all boxes that are crossed
by both defects. This list is never empty. One of these boxes is then chosen
randomly with uniform probability and flipped. The type of flip is the
one depicted in Figure~\ref{fig:oneBoxFlip} (c) and the resulting configuration 
is also without loops. However, one of the defects enters from
and exits at the top while the other goes bottom to bottom. By repeating
the algorithm we get a new configuration with no loops and two defects 
crossing the lattice from top to bottom.

How does one get a starting configuration without closed loops? We fill the lattice
randomly. The result is a configuration with (in general) many loops and two
defects. We then repeat the following step. We list boxes that are crossed
only once by either defect. If there is a closed loop, this list is non empty.
We then flip one of these boxes, thereby changing the left configuration of 
Figure~\ref{fig:oneBoxFlip} (b) to its right companion, and start over. At the
end we can check that the defects cross from top to bottom and, if necessary,
correct by (half) the above algorithm.

\section{Initial thermalization and independence of measurements}

Let us call an MC step the trial of flipping $20$ boxes at once. 
We have established experimentally the number of MC steps necessary to
assure that measurements of fractal dimension of defects or of other masses are
statistically independent. As is well-known, this number of steps strongly depends
on the model. For example, the $4$-Potts model requires a significantly larger number of
steps than the Ising model. For large lattices, we use the following number of MC
steps between measurements
\be
 \frac{H\times V}{8\times 20}d(\beta)
\ee
where $d(\beta)$ is a factor dependent on the model: $d=1$ for ${\cal LM}(3,5)$,
$d=2$ for the logarithmic Ising model ${\cal LM}(3,4)$ and $d=4$ for the 
logarithmic tricritical Ising model ${\cal LM}(4,5)$. 
Hence, for ${\cal LM}(3,5)$, we
try to flip one eighth of the boxes (by trials of $20$ boxes) between measurements.
This number goes up to one half for ${\cal LM}(4,5)$. 
A finer analysis is under way and will be published elsewhere. 
It confirms that these numbers are fine, probably on the zealous side.

The initial random configuration is thermalized by $1000$ times the number
of MC steps just listed. The initial thermalization is therefore four times
longer for ${\cal LM}(4,5)$ than for ${\cal M}(3,5)$.
Again, we think that this is satisfactory, if not ``too prudent'', but this is harder to assess.
We have experimented with the (spin) Ising model and compared it with the $\beta=\sqrt2$
loop gas as follows. First, we use an ordinary Ising spin lattice wrapped
on a cylinder at the critical temperature. There are four times as many spins
along the cylinder than along the
circular sections. We measure the two-point spin
correlation function on the section in the middle of the cylinder using many
samples of $10^3$ configurations (chosen to be independent). We then perform the
same experiment with the corresponding loop gas. We know that a loop configuration can
be interpreted as an FK configuration and therefore that a spin
configuration can be constructed from it. We repeat several times the
following experiment: from a random configuration, we apply $N$ MC steps and
then proceed to measure the spin correlation function with $10^3$ configurations,
each separated by the number of MC steps assuring statistical independence. By
comparing the distributions of the correlation function for the spin model,
on one hand, and the loop gas on the other hand, we find that it is 
sufficient to use the $N$ stated earlier, that is $N=1000$, times the number
of MC steps between measurements.

For critical dense polymers ($\beta=0$), no initial thermalization is necessary. We flip
$H\times V/8$ boxes between measurements. 

As the boxes are chosen randomly (for 
critical dense polymers and the other models), the number of boxes actually flipped
might be smaller since, for example, the same box
can be flipped twice between two measurements.
Moreover, if a box is chosen more than once among the $20$ being flipped, it will be flipped
only once and the total number of boxes flipped during this Metropolis step will be
smaller than $20$. This assures that the number of boxes flipped at each step
can be of either parity.

We use L'\'Ecuyer and Tezuka's pseudo-random number generator~\cite{LEcuyer} in the simulations.

%
%


\end{document}